\documentclass[%
 reprint,
 amsmath,amssymb,
 aps,
]{revtex4-1}

\usepackage{graphicx}% Include figure files
\usepackage{dcolumn}% Align table columns on decimal point
\usepackage{bm}% bold math
\usepackage{color}
\usepackage{amsthm}
\usepackage{subfig}

\newcommand{\Author}[1]{{#1} \textit{et al.}}

\newtheorem{definition}{Definition}
\newtheorem{theorem}{Theorem}

\begin{document}

\preprint{APS/123-QED}

\title{Community Detection Using Multilayer Edge Mixture Model}

\author{Han~Zhang}
\email{zh950713@gmail.com}
\author{Chang-Dong Wang}
\email{changdongwang@hotmail.com}
\author{Jian-Huang~Lai}
\email{stsljh@mail.sysu.edu.cn}
\affiliation{
School of Data and Computer Science, Sun Yat-sen University,  Guangzhou, P. R. China.
}
\author{Philip S. Yu,}%
 \email{psyu@cs.uic.edu}
\affiliation{%
University of Illinois at Chicago, Chicago, IL 60607, USA.
}%

\date{\today}% It is always \today, today,

\begin{abstract}
A wide range of complex systems can be modeled as networks with corresponding constraints on the edges and nodes, which have been extensively studied in recent years.
Nowadays, with the progress of information technology, systems that contain the information collected from multiple perspectives have been generated.
The conventional models designed for single perspective networks fail to depict the diverse topological properties of such systems, so multilayer network models aiming at describing the structure of these networks emerge.
As a major concern in network science, decomposing the networks into \textit{communities}, which usually refers to closely interconnected node groups, extracts valuable information about the structure and interactions of the network.
Unlike the contention of dozens of models and methods in conventional single-layer networks, methods aiming at discovering the communities in the multilayer networks are still limited. In order to help explore the community structure in multilayer networks, we propose the multilayer edge mixture model, which explores a relatively general form of a community structure evaluator from an edge combination view.
As an example, we demonstrate that the multilayer modularity and stochastic blockmodels can be derived from the proposed model.
We also explore the decomposition of community structure evaluators with specific forms to the multilayer edge mixture model representation, which turns out to reveal some new interpretation of the evaluators.
The flexibility and performance on different networks of the proposed model are illustrated with applications on a series of benchmark networks.
\end{abstract}

\pacs{Valid PACS appear here}% PACS, the Physics and Astronomy
\maketitle

%\tableofcontents

\section{Introduction}
\label{sec:introduction}
%*** Background Introduction ***
%1. Introduce the networks, multilayer networks and their applications (the rapid increase of the empirical data of multilayer networks).
Networks have been widely used in characterizing complex systems in various areas such as transportation networks, electrical networks, social networks, and biological networks, \textit{etc.}~\cite{strogatz:2001exploring,newman:2010networks,wasserman:1994social,girvan:2002community}. Traditionally, a network is represented as a graph where the nodes represent individuals of the network and the presence of an edge between a pair of nodes indicates their connection~\cite{bollobas:1998modern}. In a more complex scenario, a variety of attributes of the edges are explored, which lead to directed graphs~\cite{newman:2010networks,bang:2008:digraphs}, weighted graphs~\cite{newman:2004analysis,barrat:2004architecture}, signed graphs~\cite{doreian:2009partitioning,yang:2007community} and so on. Although these graphs have successfully depicted a wide range of network systems, they fail to comprehensively construe the network structure when the edges are distinguished into multiple types or the network is temporal (i.e. the edges of the network vary over time)~\cite{verbrugge:1979multiplexity,szell:2010multirelational,rocklin:2013clustering,holme:2012temporal}. For instance, consider a phone call temporal network where the calling links are recorded for a series of time. This network consists of multiple time slices of these calling states. Notice that if there is an edge between two individuals in two successive time slices, they may have a long phone call that last through the two time points or they may have two independent calls. To distinguish such circumstances, the interdependency of the slices should be taken into account. In recent years, such networks with multiple interdependent ``layers" which are represented by different graphs that describe the network from different perspectives (also called ``relations", ``edge colors", ``node colors" in related works) have sprung rapidly especially in transportation, gene and on-line networks~\cite{cardillo:2013emergence,li:2011integrative,szell:2010multirelational}, from which we always obtain more detailed and exhaustive understanding of the system. Other terminologies in this literature such as ``multigraph", ``multiplex network", ``multirelational network", ``multislice network", ``multilevel network", ``network of network" and ``temporal network" always refer to a similar network structure~\cite{kivela:2014multilayer}. In this work we will refer to such network as a \textbf{multilayer network} to avoid confusion.

\begin{figure}[!htb]
\centerline{
\includegraphics[scale=0.5]{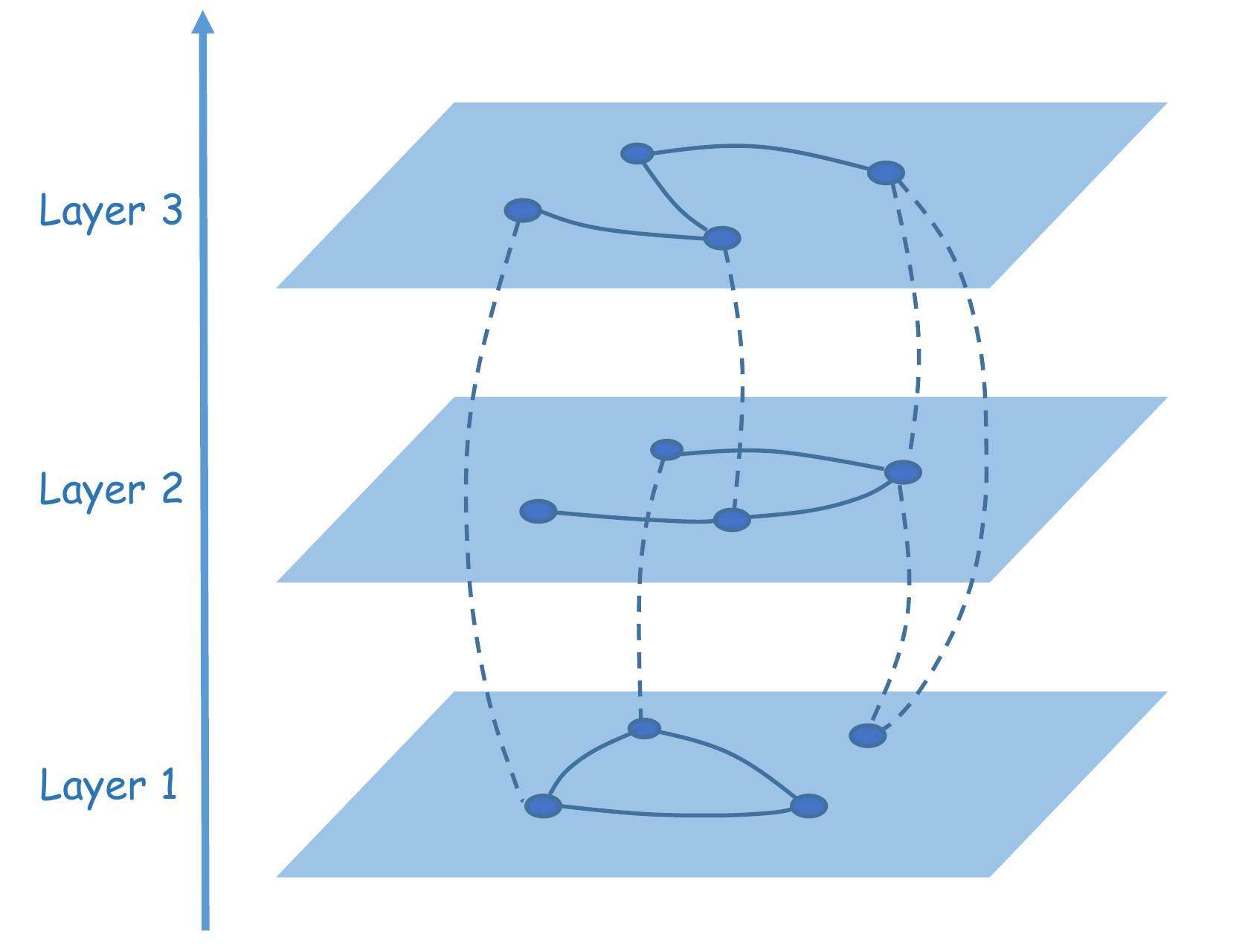}
}
\caption{The multilayer network model considered in this paper. All the layers share the same node set and are represented by a simple graph. The layers are coupled by the ``couplings" between one node and its copy in other layers, plotted as the dotted lines. It is also assumed that there is no edges connecting different global nodes in different layers.}
\label{fig:model}
\end{figure}

During the process of exploring the multilayer networks, different network representations have been explored, see Ref.~\cite{kivela:2014multilayer} for a detailed discussion.
The adopted model in this paper is the one considered by \Author{Mucha}~\cite{mucha:2010community}, as illustrated in \figurename~\ref{fig:model}.
This model assumes that the layers share the same node set and are linked by the \textit{couplings} between the node in one layer and its copies in other layers, through which the interdependency of the layers are reflected.
Recall the example of phone call network.
The network then can be represented as a multilayer network, where individuals are represented as the nodes and phone calls are represented as the edges.
Each time slice corresponds to a layer, where couplings between them indicate the continuity of the call.
Note that here the couplings appear in pairs, corresponding to the two individuals having a long call.
In this way, a wide range of complex systems can be finely depicted without loss of information about their interactions.

A major interest of network study is focused on the community structure (or assortative mixing, mesoscopic structure in the literature~\cite{newman:2002assortative,holme:2012temporal}) which is believed to reveal a coarse-grained structure of the network.
A \textit{community} usually refers to a group of nodes that are compactly connected with each other and sparsely connected with the nodes outside the group, albeit there is actually no universal definition of a ``community", since the concept of a community is application-dependent~\cite{fortunato:2010community}. Therefore, based on different assumption on the features of the communities, many objective functions have been developed to qualify the detected community structure: edge beweenness ~\cite{newman:2004finding}, edge clustering coefficient~\cite{radicchi:2004defining}, modularity~\cite{newman:2006modularity}, spin model~\cite{reichardt:2006statistical}, role model~\cite{reichardt:2007role}, infomation-theoretic methods~\cite{rosvall:2007information,rosvall:2008maps}, \textit{etc}. See Ref.~\cite{fortunato:2010community} for a review. Most existing community detection methods choose a specific quality function and optimize it. The partition with the largest functional value is recognized as the optimal community assignment.

Due to the complexity of the multilayer network structure, there is relatively a lack of approaches suitable for community detection in multilayer networks compared with the single-layer case. So far as we are concerned, existing evaluations for community structure in multilayer networks include modularity~\cite{mucha:2010community}, stochastic blockmodels (SBMs)~\cite{valles:2014multilayer,peixoto:2015inferring}, information-theoretic method~\cite{de2015:identifying} and other methods which are not focused in this paper ~\cite{battiston:2013metrics,brodka:2010method,de:2013centrality,lambiotte:2012ranking}.
By comparing the difference of the Laplacian operator with the steady state of the dynamic, the multilayer modularity describes the stability of a partition~\cite{mucha:2010community}. % and provides a node-level view of the community structure~\cite{mucha:2010community}.
%The multilayer stochastic blockmodels and information-theoretic methods describe the community structure on a community level.
The SBMs adopt a probabilistic perspective on the community structure, by introducing a specific probability distribution to the edges between communities and making inferences given the observed network adjacency as evidence~\cite{peixoto:2015inferring}.
Unlike the previous two types of methods, the information-theoretic method used in the multilayer community detection optimizes a multilayer map equation that evaluates the description length of a random walk process on the network, based on the assumption that reasonable community structure will greatly compress the information needed to represent a flow on the network~\cite{de2015:identifying}.
A more detailed  review on these approaches is given later in section \ref{sec:relatedwork}.

In order to help researchers better exploit the multilayer community structure, in this paper, inspired by the rewarding scheme proposed by Reichardt and Bornholdt~\cite{reichardt:2006statistical}, we propose a multilayer edge mixture model (MEMM) based on a linear combination of the edge contributions, which explores a relatively general form of the community structure descriptor from an edge-mixture view.
The proposed MEMM is positioned as a hyper model that provides a new interpretation of a variety of existing multilayer community structure evaluators and paves the way for the derivation of new quality functions. In particular, a specific quality function for community structure in multilayer networks can be represented as a mixture of eight types of edges using MEMM, which reveals the preference of such evaluators.
We can also utilize MEMM to develop new quality functions that reflect the preference of the definition of a ``community" in a real-world application by choosing appropriate values for the hyper parameters.
As an example, we demonstrate the derivation of modularity and SBM in the multilayer case from MEMM and decompose evaluators with specific forms to the MEMM representation.
We point out that the multilayer information-theoretical method (Infomap) has not been derived from MEMM since it takes an entirely different understanding of the community structure, albeit we will give a brief review about it in section \ref{sec:relatedwork} to appreciate its contribution to the community detection problem in multilayer networks.

The rest of this paper will be arranged as follows. We briefly review the related work in the literature in section \ref{sec:relatedwork}. We will introduce the multilayer edge mixing  model in section \ref{sec:ourmetric}. The experimental results are reported in section \ref{sec:experiments}. We conclude this paper in section \ref{sec:conclusion}.

\section{Background}
\label{sec:relatedwork}
Currently, the major efforts made to the multilayer community detection consist of modularity-based methods~\cite{mucha:2010community,bazzi:2014community}, stochastic blockmodels (SBMs)~\cite{peixoto:2015inferring} and information-theoretic methods~\cite{de2015:identifying}.
Although not the focus of this paper, other methods also provide considerable solutions to the problem, for example the edge centrality, clustering coefficient and methods based on dynamic processes~\cite{brodka:2010method,lambiotte:2012ranking,battiston:2013metrics,de:2013centrality}.

%explain modularity
\textit{Modularity} is a widely adopted evaluation for community structure in single-layer networks~\cite{newman:2004finding,clauset:2004finding,newman:2006modularity,newman:2010networks}.
The original definition of modularity is the edge difference between current network and a \textit{null model}, which is a rewired network whose edges are uniformly redistributed while maintaining the degree distribution.
Modularity reflects the cohesion of nodes within a community, so by optimizing global modularity value one can unfold communities with dense intra-community edges~\cite{newman:2006modularity}.
Recently, \Author{Mucha} extended the single-layer modularity to the multilayer case by utilizing Laplacian dynamics defined on a multilayer network, which measures the stability of a community by comparing the current one with the static distribution, which is proved to have the same form with the null model~\cite{mucha:2010community,lambiotte:2008laplacian}.
%It is worth noticing that the static solution of random walk starting from a specific node in this case depends merely on the local structure of the node itself (the degree) and the scale of the network (total number of edges). Thus, comparing the Laplacian operator (local adjacency) with the static solution will lead to a local feature. Therefore, the multilayer modularity provides a node-level evaluation for the community structure. The multilayer modularity is of great contribution due to the fact that it combines the layer on a model level for the first time and is adopted in a wide range of areas~\cite{szell:2010multirelational,bassett:2011dynamic,chiu:2011unifying}.

In the literature, a \textit{stochastic blockmodel} (SBM) usually refers to the model that describes the link structure between node groups of the network~\cite{holland:1983stochastic,karrer:2011stochastic}.
SBMs regard the community as a group of nodes with similar linking probability.
By modeling the probability of having edges between groups, providing the graph adjacency as evidence, we are able to obtain the community assignment~\cite{karrer:2011stochastic} and infer the missing edges~\cite{guimera:2009missing}.
So far the SBMs in multilayer networks adopt the idea of ``independent layers" to represent the multilayer networks, i.e. ignoring the couplings between layers~\cite{peixoto:2015inferring,barbillon:2015stochastic}.
One either aggregates (or ``collapse" in the related work) the layers to generate a single-layer network, or trains the blockmodel in each layer before inferring the global community structure~\cite{barbillon:2015stochastic,peixoto:2015inferring,stanley:2015clustering,taylor:2015enhanced,paul:2015community}.
The multilayer SBMs provide a promising probabilistic approach to solve community detection, with flexible choices of probability distribution of generating edges dealing with various network structures.

\textit{Information-theoretic methods} take another perspective of the issue of community structure. Such models assume that by utilizing the community structure, one can compress the information needed to describe the whole network~\cite{rosvall:2007information,rosvall:2008maps}. This process can be modeled as the signal reconstruction problem with the limited signal channel capacity~\cite{rosvall:2007information} or the coding length minimization problem to represent an infinite random walk dynamic~\cite{rosvall:2008maps}. The former problem is tackled by comparing the mutual information of the original and reconstructed structures of the network, while the latter one minimizes the map equation (the essence is Shannon entropy) to reduce the average description length. Based on the work of \Author{Rosvall}~\cite{rosvall:2008maps}, \Author{De Domenico} proposed a multilayer information-theoretic method called \textit{informap} using the multilayer map equation~\cite{de2015:identifying}. By taking advantage of a two-level code scheme,
%: giving unique names to the communities only and recycling the short code names among the communities,
map equation is represented as a sum of Shannon entropy of looking up the reference code books of each community.
Unlike SBMs and modularity, informap is based on a ``flow" defined in the network, from which we can extract the frequency of looking up the code books.
The adjacency merely contributes to constructing the transition probability matrix.
To be more specific, informap is a function of the reference probability (i.e. $q_{\boldsymbol{\iota}\curvearrowleft}$ and $p_{i\in{\boldsymbol{\iota}}}$ in Ref.~\cite{de2015:identifying}), which is the sum of transition probabilities (i.e. $p_{ij}^{\alpha\beta}$).
The transition probability is a function of the adjacency, finally.
So this work is much different from the discussed models in this paper.

In a nutshell, despite of the hot research interests,  methods for community detection in multilayer networks are still limited due to the complicated network frame. To help construct evaluations for multilayer community structure, in this work, we propose a multilayer edge mixture model (MEMM) to explore a relatively general form
 of the quality function for the multilayer community structure from an edge mixture view and this will also be useful for related work on the subject.
\section{The Multilayer Edge Mixture Model}
\label{sec:ourmetric}
%Introduce our modularity representation from the Hamiltonian ($ansatz$)
In Ref.~\cite{reichardt:2006statistical}, Reichardt and Bornholdt proposed a rewarding scheme of edges to describe a general quality function for community structure in single-layer networks: (i) rewarding existing edges within a community, (ii) penalizing non-existing edges within a community, (iii) penalizing existing edges between two communities and (iv) rewarding non-existing edges between two communities. The general quality function they considered takes the form as follows:
\begin{equation}
\label{eq:singlelayeransatz}
\begin{aligned}
\mathcal{H}(g) &= -\sum_{i\neq{j}}a_{ij}\underbrace{A_{ij}\delta(g_i, g_j)}_{\text{Internal existing edges}}\\
&+ \sum_{i\neq{j}}b_{ij}\underbrace{(1-A_{ij})\delta(g_i, g_j)}_{\text{Internal non-existing edges}}\\
&+\sum_{i\neq{j}}c_{ij}\underbrace{A_{ij}\big[1-\delta(g_i, g_j)\big]}_{\text{External existing edges}}\\
&-\sum_{i\neq{j}}d_{ij}\underbrace{(1-A_{ij})\big[1-\delta(g_i, g_j)\big]}_{\text{External non-existing edges}},
\end{aligned}
\end{equation}
where $A_{ij}$ is the edge strength of nodes $i$ and $j$, $g_i$ indicates the label of the community that node $i$ belongs to, and $\{a, b, c, d\}$ is the set of free hyper parameters. The delta function $\delta$ is the Kronecker delta. A lower $\mathcal{H}(g)$ value indicates a better partition.
Despite great success, such rewarding scheme is improper for some networks, e.g. bipartite networks, where edges are expected to distribute between the communities rather than within them.
Actually, in such networks the absent edges should be encouraged instead.
This implies that using the signs to assert the contribution type is too rigid to extend to other network types.
Eq. \eqref{eq:singlelayeransatz} is similar to the evaluation of the ``role model" proposed by Reichardt and White ~\cite{reichardt:2007role}, which assumes that edges are only allowed between some pairs of communities, recognized as ``intimate communities", and are banned between other community pairs.
This evaluator encourages two kinds of edges --- existing edges between intimate communities and non-existing edges between non-intimate communities:
\begin{equation}
\label{eq:rolemodel}
  \mathcal{Q}^B(\{\sigma\}) = \frac{1}{M}\bigg(\sum_{i \neq{j}}a_{ij}A_{ij}B_{\sigma_{i}\sigma_{j}} + b_{ij}(1-A_{ij})(1-B_{\sigma_{i}\sigma_{j}})\bigg),
\end{equation}
where $M$ is the total edge weight and matrix $\mathbf{B}$ records whether the communities are intimate. If matrix $\mathbf{B}$ takes a diagonal form, then Eq. \eqref{eq:rolemodel} is equivalent to Eq. \eqref{eq:singlelayeransatz} without considering the punishment terms (the two types of edges with positive contribution).

Rather than directly finding the community assignment for each node, the above two models first introduce ``hyper parameters" and determine their value according to the definition of the ``community", to obtain an evaluator of the community structure. Then the evaluator describes the quality of the detected community structure and guides us to the optimal community assignment. We call a model taking such a two-phase approach to determine the community assignment a \textit{hyper model} to distinguish them from conventional models in the literature, which usually refers to a specific evaluator. Due to the flexibility of the selection of hyper parameters, hyper models are able to adapt to different circumstances. As an example, \Author{Reichardt} derived the Hamiltonian of a spin glass and its equivalence with modularity~\cite{reichardt:2006statistical}.

Inspired by the previous work, in multilayer networks, we propose the \textit{multilayer edge mixture model} (MEMM), which is a multilayer hyper model that enables couplings connecting pairs of layers and introduces the probabilities on the links:
\begin{equation}
\label{eq:MEMM}
\begin{aligned}
&\mathcal{M}(\upsilon)= \sum_{i\neq{j}, s}\underbrace{\lambda(\{a\})a_{ijs}A_{ijs}P(\upsilon_{is}, \upsilon_{js})}_{\text{Internal existing edges}}\\
&~~+\sum_{i\neq{j}, s}\underbrace{\lambda(\{b\})b_{ijs}(1-A_{ijs})P(\upsilon_{is}, \upsilon_{js})}_{\text{Internal non-existing edges}} \\
&~~+\sum_{i\neq{j}, s}\underbrace{\lambda(\{c\})c_{ijs}A_{ijs}[1-P(\upsilon_{is}, \upsilon_{js})]}_{\text{External existing edges}} \\
&~~+\sum_{i\neq{j}, s}\underbrace{\lambda(\{d\})d_{ijs}(1-A_{ijs})[1-P(\upsilon_{is}, \upsilon_{js})]}_{\text{External non-existing edges}} \\
&~~+\sum_{s\neq{r}, i}\underbrace{\lambda(\{e\})e_{isr}C_{isr}P(\upsilon_{is}, \upsilon_{ir})}_{\text{Internal existing couplings}} \\
&~~+\sum_{s\neq{r}, i}\underbrace{\lambda(\{f\})f_{isr}(1-C_{isr})P(\upsilon_{is}, \upsilon_{ir})}_{\text{Internal non-existing couplings}} \\
&~~+\sum_{s\neq{r}, i}\underbrace{\lambda(\{g\})g_{isr}C_{isr}[1-P(\upsilon_{is}, \upsilon_{ir})]}_{\text{External existing couplings}} \\
&~~+\sum_{s\neq{r}, i}\underbrace{\lambda(\{h\})h_{isr}(1-C_{isr})[1-P(\upsilon_{is}, \upsilon_{ir})]}_{\text{External non-existing couplings}},
\end{aligned}
\end{equation}
 where $s$ and $r$ denote layers, node $is$ means node $i$ in layer $s$, and matrix $\mathbf{A}$, $\mathbf{C}$ and $\boldsymbol{\upsilon}$ denote the within-layer adjacency, between-layer adjacency and the community label matrix, respectively.
The hyper parameter set $\{a, b, c, d, e, f, g, h\}$ is the mixture coefficients that control the contribution of the corresponding edge type.
The $P(\upsilon_{is}, \upsilon_{jr})$ can be interpreted as the indicator of how evident node $is$ and node $jr$ belong to the same community or the probability of having edges between the communities to which node $is$ and $jr$ belong.
By using a probability representation rather than the $\delta$ function like Eq. \eqref{eq:singlelayeransatz}, MEMM enables a ``fuzzy" partition of the network in some cases, as we do not assert that two nodes belong to either the same or different communities.
We can easily revert to a ``hard'' division by taking the probability matrix $\mathbf{P}$ as diagonal and rounding the entries to binary values.
The $\lambda(\{\omega\})$ function is an indicator of whether a larger or smaller value of $\omega$ leads to a greater contribution:
\begin{equation}
  \lambda(\{\omega\}) =
  \begin{cases}
    +1 & \text{if a larger value of $\omega$ leads}\\
    & \text{to a greater contribution,}\\
    -1 & \text{otherwise.}
  \end{cases}
\end{equation}
Unlike the treatment in Eq. \eqref{eq:singlelayeransatz}, the sign of each term does not indicate the encouragement (or punishment) in MEMM.
Instead, we introduce the $\lambda$ function so that the hyper parameters can take values with arbitrary signs.
Meanwhile, using $\lambda$ function rather than the signs of the terms makes MEMM more flexible to deal with different network structure (e.g. the bipartite network problem mentioned above is now well addressed, as will be confirmed in experiments.).

Note that MEMM is not based on the exact definition of the community or the probability distribution. It just encourages the presence (or absence) of the existing (or non-existing) of the edges (or couplings) within (or between) communities, weighted by the hyper parameters $\{a, b, c, d, e, f, g, h\}$ and the corresponding linking probabilities.
%This should always hold since the community structure is just inferred by the edge distribution, given the node set.
Moreover, MEMM formally describes the quality of community structure as a mixture of eight types of edges.
By representing a quality function in this way, the preferences of this quality function can be reflected by the contribution weights, e.g. whether the cohesion or adhesion of the community structure is more concerned.
In fact, the \textit{edge betweenness}~\cite{newman:2004finding} and \textit{edge clustering coefficient}~\cite{radicchi:2004defining} are just doing the same job --- finding the contribution of different edges.
The difference is, after deciding the contribution of each edge, they utilize the edge betweenness value or edge clustering coefficient of each edge separately, to distinguish the edges between communities and within communities.
The edges are not distinguished until their contributions are calculated.
In MEMM, the edges are distinguished in advance so that we can determine their contributions appropriately to flexibly construct a desired evaluator.
We also point out that, since the proposed objective function is a linear combination, the decomposition of a specific function is not unique, which enables it to adapt to different scenarios, as will be discussed in section~\ref{sec:decomposition}.

Before studying the community assignments, we have eight free hyper parameter sets $\{a, b, c, d, e, f, g, h\}$ to determine, which control the contribution of eight types of edges (couplings). Two strategies can be used to choose the parameters, i.e. to take them as fixed values or specific functions of the network structure.
We may assign fixed values to the hyper parameters once we have determined the contribution proportion of the edges (couplings) in the network. Nevertheless, if the observed network changes, we have to reassign the parameters.
In order to enable the automatic adjustment according to the network structure, the hyper parameters are assigned with values dependent on the linking adjacency.
Later in this section we will demonstrate the derivation of modularity and SBM as an example.

It is worth noticing that if the hyper parameters have different domains (for example,  $\{a\}$ can take a sufficiently larger magnitude than other parameters), the model may degenerate into a simpler one. This degenerated model just consists of the preferred terms, since even the largest values other parameters can take will be overpowered by the terms with weights $\{a\}$.
As a result, we should choose the hyper parameters for MEMM in a manner that guarantees distinguishing the contribution of different edges and couplings while avoiding being dominated by specific terms.
This requires the choice of the parameters to be \textit{discriminative}: i) if we choose two parameters to take the largest value they can take, equal attention is paid to both terms; ii) if we need to emphasize a specific edge type from the others, we just make that term take a relatively large value in its domain and others take a relatively small value in their own codomains. Note that the terms with smaller weights still have the ``potential" to take effect. Accordingly, we have the definition on discriminative as follows.
\begin{definition}
\label{def:1}
 A hyper parameter set $\{a, b, c, d, e, f, g, h\}$ of MEMM is discriminative if
\begin{align}
\nonumber
&\frac{\max_{x_1}\lambda(\{\omega_1\})\omega_1(x_1)}{\max_{x_2}\lambda(\{\omega_2\})\omega_2(x_2)} \rightarrow 1,\\
\nonumber&\forall{\omega_1, \omega_2}\in{\{a, b, c, d, e, f, g, h\}}.
\end{align}
\end{definition}
Definition \ref{def:1} states that the hyper parameters should have the same contribution when they reach the respective maxima (with $\lambda$).
This means by selecting a discriminative hyper parameter set, the terms of MEMM has equal potential to affect the global functional value.
As a basis of the derivations later in this paper, we claim a possible discriminative choice of the parameters.
\begin{theorem}
\label{thm:1}
Suppose $F_1$ and $F_2$ are continuous monotonically increasing functions and satisfy $F_1(U) = F_2(L)$. Then the hyper parameter set is discriminative if
$$\forall x_i \in [L, U],$$
$$\forall \omega_i \in \{a, b, c, d, e, f, g, h\},$$
$$
\omega_i =
\begin{cases}
F_1(x_i), &\text{if $\lambda(\omega_i) > 0$,}\\
F_2(x_i), &\text{otherwise.}
\end{cases}
$$
\begin{proof}
Since the functions $F_1$, $F_2$ are continuous and monotonically increasing, the maxima of $\lambda(\{\omega_i\})$ is reached at $F_1(U)$ if $\lambda(\{\omega_i\}) > 0$ or $F_2(L)$ otherwise. According to the assumption that $F_1(U) = F_2(L)$, the maxima of the two hyper parameters are equal. We can verify that for all $\omega_i \in \{a, b, c, d, e, f, g, h\}$, Definition \ref{def:1} holds.
\end{proof}
\end{theorem}

With Theorem \ref{thm:1}, we are able to adjust MEMM by introducing desired functions to the hyper parameters. Here we let the encouraged (punished) edges share the same weight function to reduce the independent terms and obtain a simpler representation.

\subsection{Deriving Modularity from MEMM}
\label{sec:deriveModularity}
\Author{Mucha} proposed the multilayer modularity based on a Laplacian dynamic defined on the multilayer networks~\cite{mucha:2010community}. They assume that a random walker tends to stay in the same community after repeated random jumping. With Theorem \ref{thm:1}, we demonstrate that MEMM provides another interpretation of multilayer modularity.

Here we  take $\lambda(\{\omega\}) = +1$ for $\omega \in \{a, d, e, h\}$, $\lambda(\{\omega\}) = -1$ for other hyper parameters and
 \begin{equation}
 \label{eq:choiceOfModularity}
 \begin{cases}
   &a_{ijs}=c_{ijs} = 1-\gamma_{s}p_{ijs}\\
   &b_{ijs}=d_{ijs}=\gamma_s p_{ijs}\\
   &e_{isr} = g_{isr} = \varsigma \\
   &f_{isr} = h_{isr} = 0,
 \end{cases}
 \end{equation}
 where $p_{ijs}$ is the null model in layer $s$, $\gamma_s$ is the corresponding resolution parameter and $\varsigma$ controls the coupling strength between the layers.
Temporarily, by ignoring the contribution of non-existing couplings here, we obtain
\begin{equation}
\label{eq:fuzzyModularity}
  \begin{aligned}
    &\mathcal{M} (\upsilon) = \sum_{i\neq{j}, s}(A_{ijs}-\gamma_sp_{ijs})[2P(\upsilon_{is}, \upsilon_{js})-1]\\
    &~~~ + \sum_{s\neq{r}, i}\varsigma C_{isr}[2P(\upsilon_{is}, \upsilon_{ir})-1] \\
    &= \underbrace{2\sum_{ijsr}\bigg[(A_{ijs}-\gamma_sp_{ijs})P(\upsilon_{is}, \upsilon_{js})+\tilde{C}_{isr}P(\upsilon_{is}, \upsilon_{ir})\bigg]}_{\text{Fuzzy multilayer modularity}}\\
    &~~~ \underbrace{-\sum_s(1-\gamma_s)m_s - \sum_im'_i}_{\text{Constant}} ,
  \end{aligned}
\end{equation}
where we utilize the fact that $\sum_{ij}A_{ijs} = \sum_{ij}p_{ijs}$ and define $\tilde{C}_{isr} = \varsigma C_{isr}$, $m_s = \sum_{ij}A_{ijs}$ and $m'_i = \sum_{sr}\tilde{C}_{isr}$ to keep the notations uncluttered.

Although beyond the scope of this paper, a ``fuzzy" modularity representation can be obtained from MEMM as a by-product, where the nodes are considered to be in the same community with a probability.
This reflects how reliable the current assignment is.
In this way, the summation of effective edges now runs throughout the network rather than merely within the communities.
If the probability matrix $\mathbf{P}$ is diagonal and the entries take only binary values, maximizing $\mathcal{M}(\upsilon)$ is equivalent to the optimization of the multilayer modularity proposed by \Author{Mucha}
\begin{equation}
\label{eq:multilayerModularity}
\begin{aligned}
\mathcal{M}(\upsilon) &=2\sum_{ijsr}\bigg[(A_{ijs}-\gamma_sp_{ijs})\delta_{sr}+\tilde{C}_{isr}\delta_{ij})\bigg]\delta(\upsilon_{is}, \upsilon_{jr})\\
    &~~~-\sum_s(1-\gamma_s)m_s - \sum_im'_i,
\end{aligned}
\end{equation}
where the $\delta$ function is the Kronecker delta.

Now let's take a closer look at the derivation.
The hyper parameter set is discriminative since both $F_1(-p_{ijs})=1-\gamma_s p_{ijs}$ and $F_2(p_{ijs})=\gamma_sp_{ijs}$ are continuous monotonically increasing functions.
The maxima of the functions are both equal to one by taking $-p_{ijs}=0$ in $F_1(-p_{ijs})$ and $p_{ijs}=1$ in $F_2(p_{ijs})$, where we assume the resolution parameter $\gamma_s$ takes 1, as in most cases.
Therefore, such choice satisfies Theorem \ref{thm:1}, and the value of $p_{ijs}$ balances the contribution of the edges.
Recall that the probability of linking in the null model of each layer is represented as $\gamma_sp_{ijs}$. This can be interpreted as the expected edge strength between node $is$ and $js$ in a randomly rewired network (in unweighted networks).
Therefore, the effective edge strength of the observed network is $A_{ijs}-\gamma_sp_{ijs} \in \{-\gamma_sp_{ijs}, 1-\gamma_sp_{ijs}\}$. We find that the effective edge strength matches the value of the parameters (with $\lambda(\{\omega\})$) we pick to derive the modularity.
This suggests that the parameters here act as the introduction of the null model.
Based on the specific choice of the null model here, it is equivalent to a reconstruction of the network adjacency.
For example, if we choose the Newman-Girvan null model $\frac{k_{is}k_{js}}{2m_s}$~\cite{newman:2006modularity}  , where $k_{is}$ denotes the degree of node $is$ within the layer and $2m_s = \sum_{i}k_{is}$, such choice of the hyper parameters is equivalent to converting the adjacency matrix $\mathbf{A}$ to the modularity matrix $\mathbf{B}$.
It is worth noticing that the choice of the hyper parameters is not unique.
Indeed, we can make $\{a, b\}$ or $\{c, d\}$ to be zero and obtain the same representation as Eq. \eqref{eq:multilayerModularity} (the scalar 2 is absorbed into $\varsigma$ for couplings) disregarding the constant terms.
This indicates that the modularity can be represented subject to either the internal or external edges within the layers, as explored in the literature ~\cite{newman:2006modularity}~\cite{djidjev:2006scalable}.
But in either way, the modularity favors communities with densely distributed efficient edges.
We can reverse the preference of modularity by changing the signs of the $\lambda$ function for desired terms.

We can also take non-existing couplings into account by setting $f_{isr} = h_{isr} = \varsigma$.
This will lead to a different coupling contribution, where $\tilde{C}_{isr}=\varsigma(2C_{isr}-1)$.
Then $\tilde{C}_{isr}$ can take negative values, which decreases the global modularity value.
We will examine the difference of these two weighting schemes for the couplings in experiments.
Rather than utilizing a dynamic process, we naturally obtain the multilayer modularity based on MEMM by comparing the edge distribution of the observed network with the null model, which returns to the original definition of modularity~\cite{newman:2006modularity}.
Similar tricks are possible to introduce a null model of the couplings by taking into account the efficient strength of the couplings.
%Actually, Eq. \eqref{eq:multilayerModularity} defines a spin glass ~\cite{reichardt:2006statistical} with the coupling strength (distinguished with the terminology we used in this paper to represent the between-layer edge) between the nodes being $J_{is, jr} = (A_{ijs}-\gamma_sp_{ijs})\delta_{sr} + \tilde{C}_{isr}\delta_{ij}$: ferromagnetic when there exists an edge or coupling between the node pair and otherwise antiferromagnetic.

\subsection{Deriving SBMs from MEMM}
Stochastic blockmodels (SBMs) are generative models that make inferences on the linking probability between communities $P(\upsilon_{is}, \upsilon_{jr})$ given the network structure as evidence~\cite{holland:1983stochastic,karrer:2011stochastic, peixoto:2015inferring}.
Existing treatments for multilayer networks using SBMs can be divided into two cases: aggregate the layers to produce a single-layer network before learning a SBM, or assume a SBM in each layer and then make decision based on the results of each layer.
However, the layers are treated independently or as components of a linear combination in this way, which ignores the peculiarity and interdependency of the layers.
What is more, in such treatments, the community assignment of nodes are assumed identical in every layers.
But the fact is that, the roles of node in different layers do not have to be (in most cases are totally not) the same.
For example, a college student may be recognized as a student in Facebook, while acting as a businessman in WeChat, in a two-layer online social network.

Although in some specific cases, it may be reasonable to use a ``collapsed" network or model the layers independently~\cite{peixoto:2015inferring}, it is highly recommended to adopt a modeling based on the multilayer network structure.
This leads to maximizing the likelihood function
\begin{equation}
\begin{aligned}
  &\mathcal{L}(\mathbf{A}, \mathbf{C}|\mathbf{P}) = \prod_{i\neq{j}, s}P(\upsilon_{is}, \upsilon_{js})^{A_{ijs}}[1-P(\upsilon_{is}, \upsilon_{js})]^{1-A_{ijs}}\\
&~~~~~  \prod_{s\neq{r}, i}P(\upsilon_{is}, \upsilon_{ir})^{C_{isr}}[1-P(\upsilon_{is}, \upsilon_{ir})]^{1-C_{isr}}.
\end{aligned}
\end{equation}
In practice, we always deal with the logarithm of the likelihood function, which has the form as follows:
\begin{equation}
\label{eq:SBMlikelihood}
  \begin{aligned}
    &\log{\mathcal{L}(\mathbf{A}, \mathbf{C}|\mathbf{P})} = \sum_{i\neq{j}, s}A_{ijs}\log P(\upsilon_{is}, \upsilon_{js})\\
    &~~~~~+\sum_{i\neq{j}, s}(1-A_{ijs})\log[1-P(\upsilon_{is}, \upsilon_{js})]\\
    &~~~~~+\sum_{s\neq{r}, i}C_{isr}\log P(\upsilon_{is}, \upsilon_{ir})\\
    &~~~~~+\sum_{s\neq{r}, i}(1-C_{isr})\log[1-P(\upsilon_{is}, \upsilon_{ir})].
  \end{aligned}
\end{equation}
From Eq. \eqref{eq:SBMlikelihood}, we notice that the couplings are naturally introduced to the model, which extends the idea of independent or collapsed layer that has been widely adopted so far.
By erasing or inserting the couplings between different layers, we are able to control the interdependency of the layers so as to approximate the idea of independent or collapsed layer.
Nevertheless, Eq. \eqref{eq:SBMlikelihood} distinguishes the node $is$ with its copies $ir$ in other layers by introducing $P(\upsilon_{is}, \upsilon_{ir})$, which means they can be assigned to different communities.
In this way, the peculiarity of layers are conserved and more knowledge about the roles of nodes in different layers is available.

The log-likelihood function of SBMs has a similar form with MEMM since it can be interpreted as a mixture of four types of edges. More specifically, by ignoring the terms with weight $\omega \in \{b, c, f, g\}$, letting $\lambda(\{\omega\}) = +1$ for the remaining terms and choosing $F_1(x)$ as $\frac{\log P}{P}$ for the corresponding linking probabilities, we have
\begin{equation}
\nonumber
\begin{cases}
  &a_{ijs} = \log{P(\upsilon_{is}, \upsilon_{js})}/P(\upsilon_{is}, \upsilon_{js})\\
  &d_{ijs} = \log{[1-P(\upsilon_{is}, \upsilon_{js})]}/[1-P(\upsilon_{is}, \upsilon_{js})]\\
  &e_{isr} = \log{P(\upsilon_{is}, \upsilon{ir})}/P(\upsilon_{is}, \upsilon{ir})\\
  &h_{isr} = \log{[1-P(\upsilon_{is}, \upsilon{ir})]}/[1-P(\upsilon_{is}, \upsilon{ir})]\\
\end{cases}
\end{equation}
It is clear that Eq. \eqref{eq:SBMlikelihood} is equivalent to MEMM.
This choice of hyper parameters satisfies Theorem \ref{thm:1} because the probabilities lie in $[0, 1]$ and function $\frac{\log{x}}{x}$ is continuous monotonically increasing in $(0, 1]$, and the maxima of these terms equals 0 when the probability equal 1.

Notice that any choice of $F_1[\omega(P)]$ as $\frac{f(P)}{P}$ is able to change the form of $P$, once the function satisfies Theorem \ref{thm:1}.
It implies that to some extent, we can flexibly adjust the form of $P$ to endow various meaning to it.
In Eq. \eqref{eq:SBMlikelihood} we take $F_1[\omega(P)] = \frac{\log P}{P}$ to encourage the edges appear in accordance with the linking probabilities.
Other forms may also bring considerable significance to the model.
For instance, if we take $\log P$ as the weight for the edges and couplings, we find an interesting representation based on the negative entropy of the linking probabilities:
\begin{equation}
\label{eq:SBMentropy}
  \begin{aligned}
    &\mathcal{M}(\upsilon) = \sum_{i\neq{j}, s}A_{ijs}P(\upsilon_{is}, \upsilon_{js})\log P(\upsilon_{is}, \upsilon_{js})\\
    &~~~~~+\sum_{i\neq{j}, s}(1-A_{ijs})[1-P(\upsilon_{is}, \upsilon_{js})]\log[1-P(\upsilon_{is}, \upsilon_{js})]\\
    &~~~~~+\sum_{s\neq{r}, i}C_{isr}P(\upsilon_{is}, \upsilon_{ir})\log P(\upsilon_{is}, \upsilon_{ir})\\
    &~~~~~+\sum_{s\neq{r}, i}(1-C_{isr})[1-P(\upsilon_{is}, \upsilon_{ir})]\log[1-P(\upsilon_{is}, \upsilon_{ir})].
  \end{aligned}
\end{equation}
This evaluator also encourages the edges to satisfy the linking probabilities. But it promisingly gives an information-theoretic interpretation, based on the specific distribution $P$ takes.
Since there is no constraint on the selection of probability distribution of $P(\upsilon_{is}, \upsilon_{jr})$, we can choose any appropriate distribution. Once the probability distribution is determined, one can further simplify the quality functions and make inferences of the community assignments.

In this section we have discussed the original form of SBM.
To obtain the community assignment using such SBM, we need to make assumptions on the number of communities and the probability distribution $\mathbf{P}$ that generates the links.
A common choice of $\mathbf{P}$ is binomial distributions --- edges are generated with fixed probabilities between two nodes in two different communities.
Then we can substitute $\mathbf{P}$ with the expectation of the binomial distributions, which is the fraction of the observed edges (couplings) and all possible edges (couplings).
The remaining problem is just counting the edges and the couplings, and find an assignment that maximizes the fraction.
Notice that such SBM ignores the degree distribution in real-world networks and needs degree correction to address this problem~\cite{karrer:2011stochastic}.
However, in this paper we just discuss the most simple case to expound the interpretation of SBM based on MEMM.

\subsection{Decomposing an Evaluator to the MEMM Form}
\label{sec:decomposition}
We have demonstrated the derivation of modularity and SBM using MEMM with different selection of the hyper parameters.
Particularly, we notice that there are more than one choice of the hyper parameters to obtain a similar representation of modularity.
This indicates that the decomposition of an evaluator to the MEMM form is not unique.
However, such decomposition still reveals the preference of the evaluator.

Here, we will concentrate on a specific kind of evaluators, the value of which is represented as the sum of a linear combination of within-layer and between-layer adjacency within every communities, which qualifies the detected communities based on the linking structures within them:
\begin{equation}
\label{eq:evaluatorExample}
\begin{aligned}
  E &= \sum_{ijsr}\bigg\{\big[k_1(\mathbf{x_1})A_{ijs} + k_2(\mathbf{x_2})\big]\delta(\upsilon_{is}, \upsilon_{jr})\\
  &~~~+\big[k_3(\mathbf{x_3})C_{ijr} + k_4(\mathbf{x_4})\big]\delta(\upsilon_{is}, \upsilon_{jr})\bigg\},
\end{aligned}
\end{equation}
where $k_x$ is an arbitrary function except for those taking the adjacency $\mathbf{A}$ and $\mathbf{C}$ directly as inputs.
Here we consider a hard partition of the community assignment as is widely adopted in the literature.

By rearranging MEMM in terms of $\mathbf{A}$ and $\mathbf{C}$, we obtain
\begin{equation}
\label{eq:MEMMdecomposition}
  \begin{aligned}
    \mathcal{M}(\upsilon) &= \sum_{ijs}\bigg[(a_{ijs}-b_{ijs}-c_{ijs}+d_{ijs})A_{ijs}\delta(\upsilon_{is}, \upsilon_{js})\\
    &~~~~~~~~~~+(b_{ijs}-d_{ijs})\delta(\upsilon_{is}, \upsilon_{js})\bigg]\\
    &~~+\sum_{isr}\bigg[(e_{isr}-f_{isr}-g_{isr}+h_{isr})C_{isr}\delta(\upsilon_{is}, \upsilon_{ir})\\
    &~~~~~~~~~~~~+(f_{isr}-h_{isr})\delta(\upsilon_{is}, \upsilon_{ir})\bigg]\\
    &~~+ constant,
  \end{aligned}
\end{equation}
where $constant$ refers to the terms that are independent of the community assignment.
Here we omit the $\lambda$ function to keep the notations uncluttered, so from here to the end of this section, the hyper parameters also contain the information of contribution type.
Notice that we constrain MEMM to take a hard partition by taking $\mathbf{P}$ as diagonal matrix and rounding the entries to binary values.

Comparing the corresponding coefficients in Eq. \eqref{eq:evaluatorExample} and Eq. \eqref{eq:MEMMdecomposition}, we obtain
\begin{equation}
\label{eq:decompositionResult}
\begin{cases}
  &a_{ijs}-c_{ijs} = k_1 + k_2\\
  &b_{ijs}-d_{ijs} = k_2 \\
  &e_{isr}-g_{isr} = k_3 + k_4\\
  &f_{isr}-h_{isr} = k_4.
\end{cases}
\end{equation}
Since there are eight free hyper parameters while we just have four equations, as listed in Eq. \eqref{eq:decompositionResult}, the decomposition is not unique.
Nonetheless, Eq. \eqref{eq:decompositionResult} points out that the contribution difference of the internal and external edges (couplings) should coincide with the coefficients.
This in turn conveys the preference of the evaluator: what kind of edges (couplings) is encouraged?

Recall the multilayer modularity function given by Eq. \eqref{eq:multilayerModularity}, according to Eq. \eqref{eq:decompositionResult}, we have
\begin{equation}
\label{eq:modularityMEMMdecomposition}
  \begin{cases}
    &a_{ijs}-c_{ijs} = 2(1-\gamma_s p_{ijs}) > 0\\
    &b_{ijs}-d_{ijs} = -2\gamma_s p_{ijs} < 0\\
    &e_{isr}-g_{isr} = 2\varsigma > 0\\
    &f_{isr}-h_{isr} = 0,
  \end{cases}
\end{equation}
which can be verified by substituting the values we discussed in section \ref{sec:deriveModularity}.
From Eq. \eqref{eq:modularityMEMMdecomposition} we know that modularity favors the edges and couplings present in the communities than between them.
It also prefers the absent edges to be distributed between the communities but do not care the absent couplings.
In this way, we succeed to obtain the same conclusion on the preference of modularity as in section \ref{sec:deriveModularity}.

We highlight that the decomposition of such evaluators is able to deepen our understanding of them from an edge mixture view, so that further modifications and improvements about these evaluators will become more targeted.

\section{Experimental Results and Discussion}
\label{sec:experiments}

The proposed MEMM has a high flexibility due to the free specification on the hyper parameters before making community assignments.
In this section, we first verify that MEMM can correctly reflect our preference of the within-layer edges by studying the performance of modularity and its modified form.
Then we discuss the coupling contribution in MEMM based on two rewarding schemes for the couplings.
Finally the impact of the coupling strength $\varsigma$ and resolution parameter $\gamma_s$ is analyzed on different network settings to show that a better choice in practice is the one that obeys Definition \ref{def:1}.
For experimental purpose, we will take modularity as the quality function and utilize a Louvain-like heuristic~\cite{mucha:2010community} for optimization, but other evaluators derived from MEMM should have similar performance, and hence the discussion about other evaluators is omitted here.

%Most of the networks involved in our experiments are bench mark networks.
%We developed multilayer benchmark networks with standard community labels and calculate the normalized mutual information (NMI)~\cite{danon:2005comparing, meilua:2007comparing} between the labels and the obtained community assignment as an indicator of the detection quality.
%We first generate multiple Lancichinetti-Fortunato-Radicchi (LFR) benchmark networks~\cite{lancichinetti:2008benchmark} and then adding the couplings at random to construct a multilayer structure.
%For bipartite network, we adopt the planted partition model~\cite{condon:2001algorithms}, which is a generative model that constructs a network whose communities are random graphs, to generate the layers before the couplings are inserted.
%We also utilize the famous Zachary Karate Network~\cite{zachary:1977information} as a layer for analyzing the impact of breaking Definition \ref{def:1}.

\subsection{Rewarding Schemes}

\begin{figure*}[!htp]
\centerline{
\subfloat[Modularity on normal multilayer network]{\includegraphics[width=0.5\linewidth]{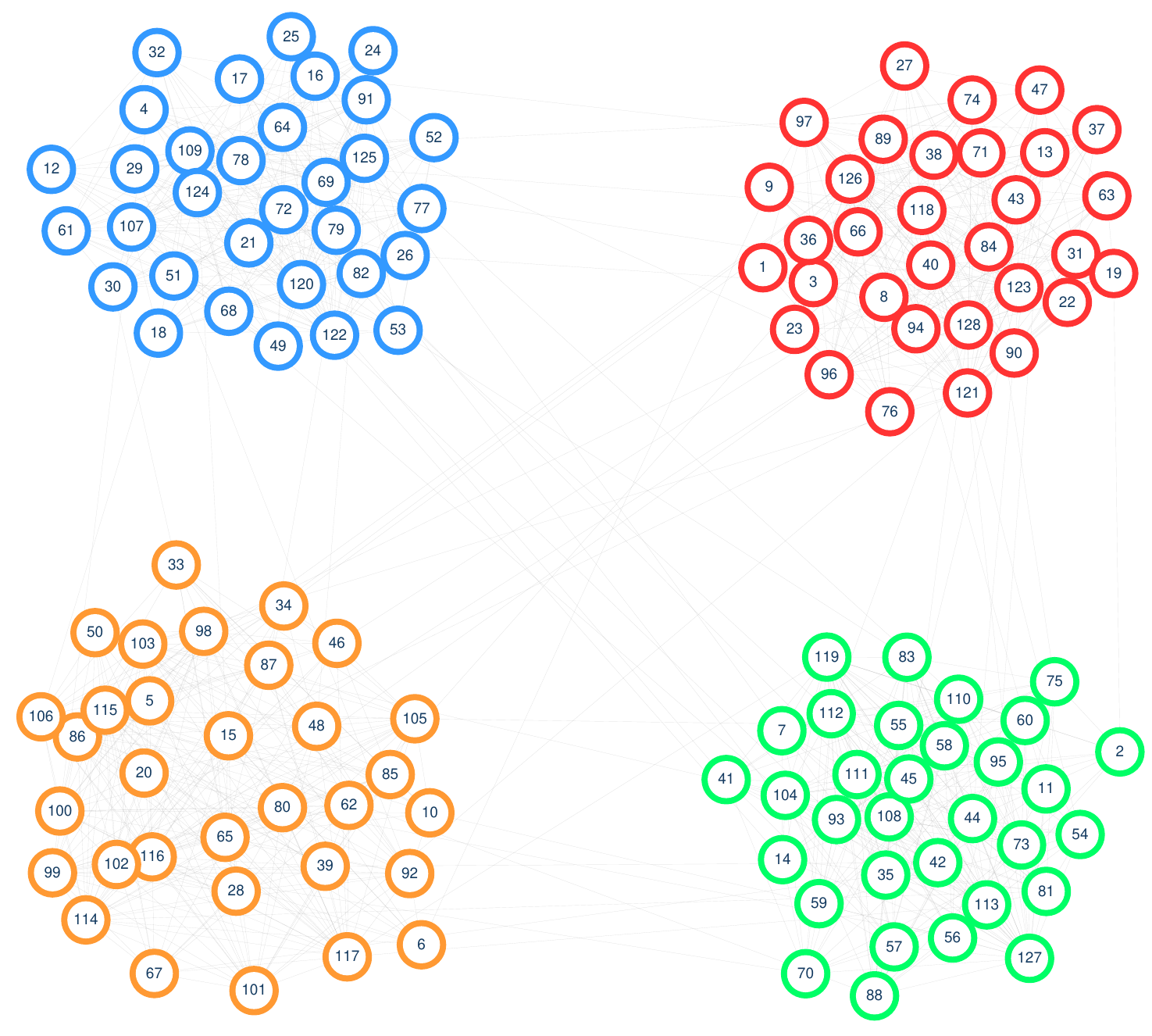}\label{fig:nornet_normal_res}}
\subfloat[Modified modularity on normal multilayer network]{\includegraphics[width=0.5\linewidth]{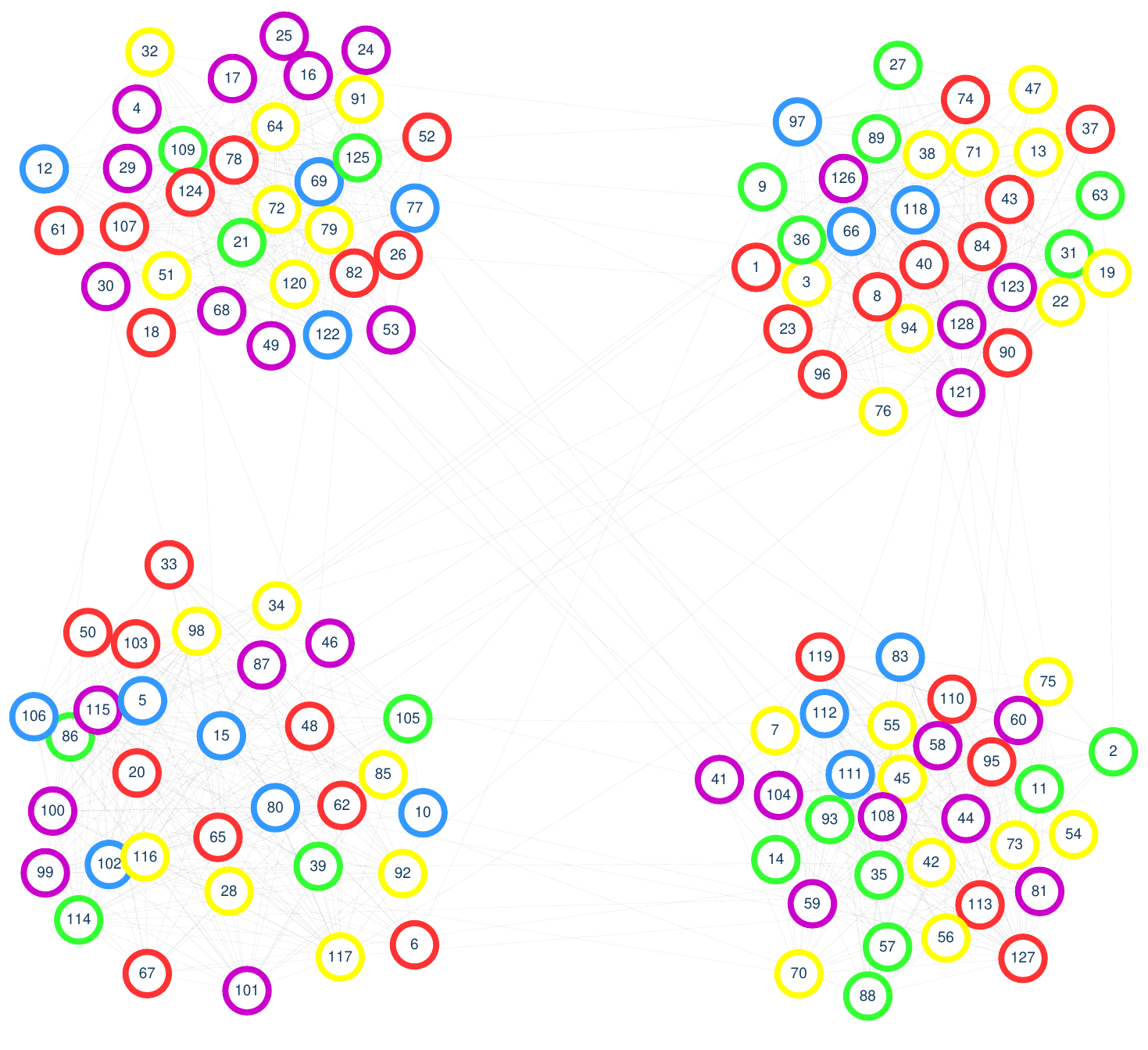}\label{fig:nornet_bipart_res}}
}
\centerline{
\subfloat[Modified modularity on bipartite multilayer network.]{\includegraphics[width=0.5\linewidth]{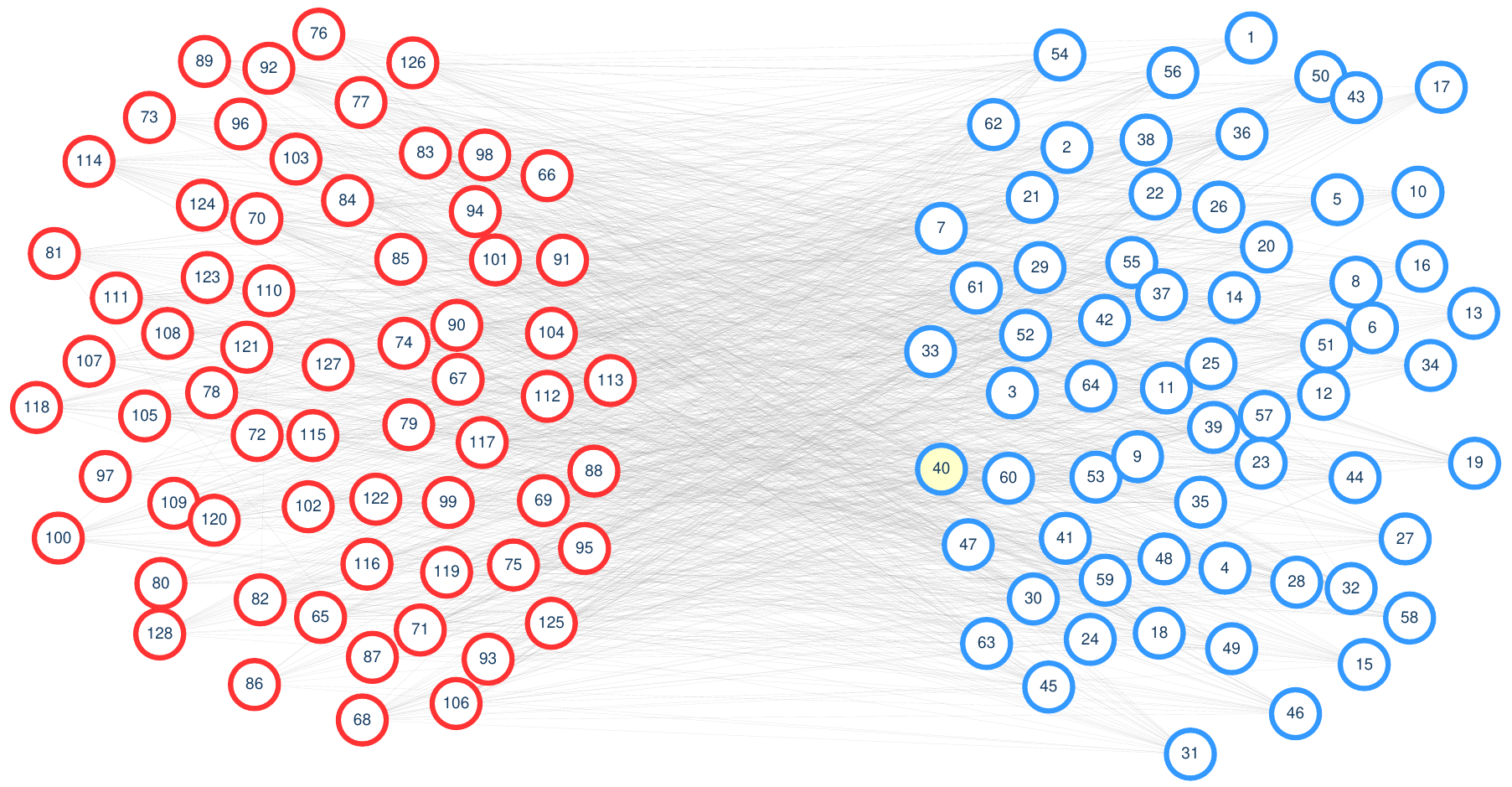}\label{fig:binet_bipart_res}}
\subfloat[Modularity on bipartite network]{\includegraphics[width=0.5\linewidth]{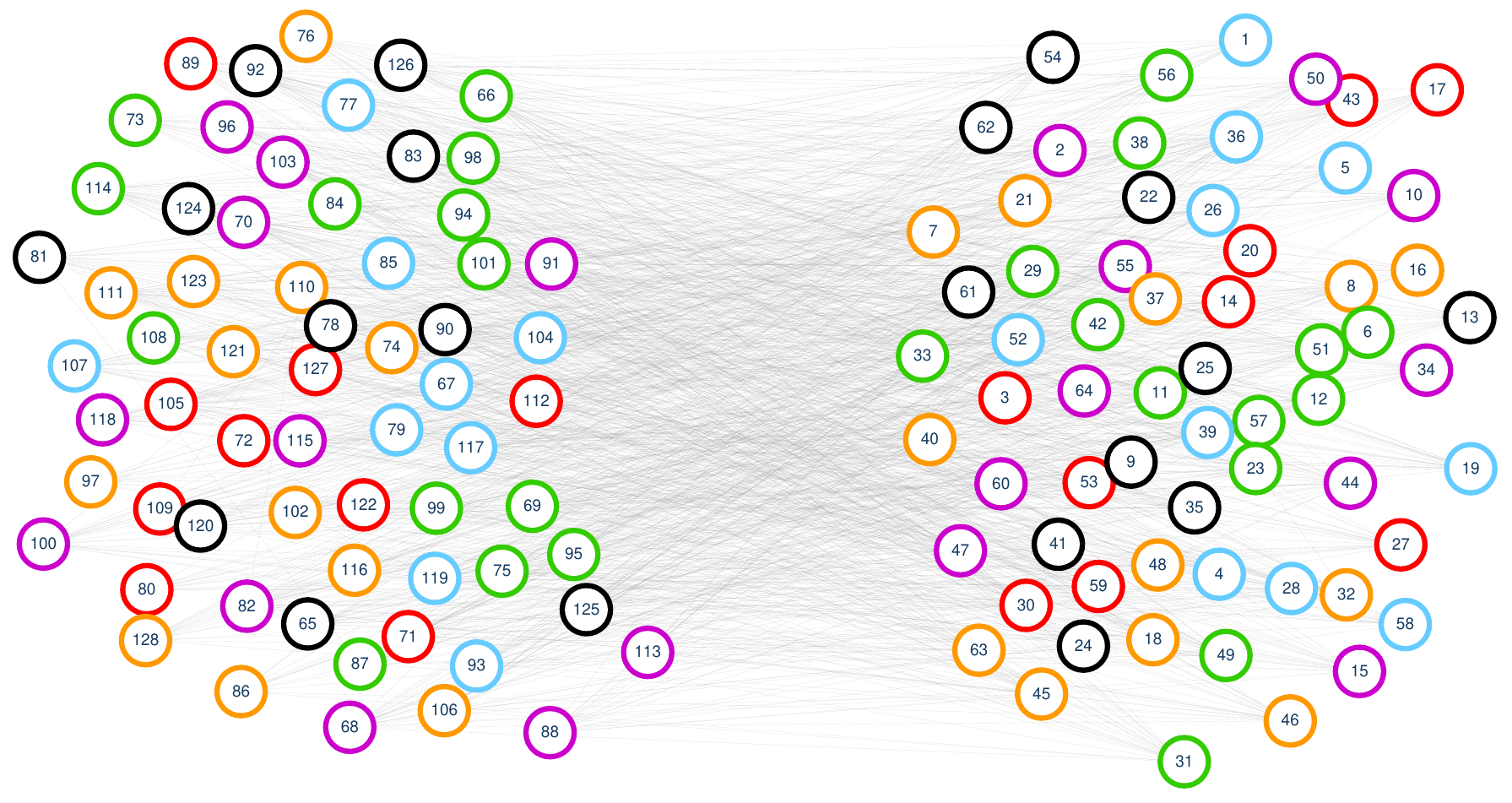}\label{fig:binet_normal_res}}
}
\caption{The community detection result of the first layers of multilayer bipartite benchmark network and normal network, respectively. The nodes are spatially partitioned into groups according to their standard community label. The obtained community assignment is depicted with different colors. }
\label{fig:differentRewardingScheme}
\end{figure*}
We begin by analyzing the influence of the rewarding scheme on the evaluator.
The proposed MEMM utilizes $\lambda$ function to distinguish the contribution type of the edges.
Edges or couplings $\omega$ with $\lambda(\{\omega\})=+1$ means such edges (couplings) are encouraged, i.e. a larger value of the corresponding hyper parameter indicates  a greater contribution.
As discussed in section \ref{sec:deriveModularity}, modularity encourages the edges to appear within communities and punishes those between the communities.
Therefore, the communities detected will have a high density of internal edges, which will lead to poor performance on bipartite (or N-partite) networks.
In fact, it has been claimed in the single-layer case that, minimizing modularity is equivalent to identifying communities in bipartite networks~\cite{newman:2006modularity}.
From an edge mixture view, we can exactly interpret this as reversing the contribution type of the edges.
In multilayer networks, if we reverse the contribution type of the within-layer edges (keeping that unchanged for the couplings), i.e. taking
\begin{equation}
  \nonumber
  \begin{cases}
    &\lambda(\{a, d, f, g\}) = -1\\
    &\lambda(\{b, c, e, h\}) = +1\\
  \end{cases}
\end{equation}
in MEMM, we obtain a modularity-like evaluator with a similar form as Eq. \eqref{eq:multilayerModularity}, except the signs of $A_{ijs}$ and $\gamma_s p_{ijs}$ are flipped.
This modified modularity should prefer the efficient edges between the communities than those within them since it has opposite preference of modularity within the layers.

To test the performance of the modularity and its modified form, we then generate two kinds of multilayer benchmark networks --- a ``bipartite" multilayer network whose layers are all bipartite networks and a ``normal" multilayer network whose layers are normal networks (the communities have dense internal edges).

 For a bipartite network, we adopt the planted partition model~\cite{condon:2001algorithms}, which is a generative model that constructs a network whose communities are random graphs, to generate the layers before the couplings are inserted.
We divide 128 nodes into two communities with equal size, and the edges between the nodes are sampled according to the corresponding linking probability, to generate a single layer.
Here we assume the edges appear between the communities with probability $p_{out} = 0.3$ while that probability of the internal edges is $p_{in} = 0.005$.
In this way, 4 layers are produced and the couplings are inserted between adjacent layers with probability $P = 1$.

For a normal network, we adopt the Lancichinetti-Fortunato-Radicchi (LFR) benchmark networks~\cite{lancichinetti:2008benchmark}, which extends the idea of the planted partition model and introduces power law distribution to the degree of nodes and the edges between the communities.
We generate 4 LFR benchmark networks as 4 layers, each of which consists of 128 nodes.
These nodes have an average within-layer degree 16 and are assigned to 4 communities with equal size.
Couplings are inserted exactly the same way as in the bipartite network.

The detection results of the first layers with different rewarding schemes on these two networks are visualized in \figurename~\ref{fig:differentRewardingScheme}. Similar results and analysis can be obtained for other layers. As expected, modularity and the modified modularity (i.e. with reversing edge contribution type) have respective advantages on the corresponding network types.
The original modularity performs well on the networks where the community members are densely connected, as shown in \figurename~\ref{fig:nornet_normal_res}, while the modified version provides a poor assignment, as shown in \figurename~\ref{fig:nornet_bipart_res}.
The rewarding scheme adopted by the original modularity encourages the efficient internal edges, which lead to communities with high cohesion.
In contrast, the modified modularity adapts to the networks where community members are sparsely connected, which allows it to uncover communities with high adhesion, as illustrated in \figurename~\ref{fig:binet_bipart_res}.
On the other hand, the original modularity shows unsatisfactory performance, as illustrated in \figurename~\ref{fig:binet_normal_res}.
Such distinguished preference reflected in MEMM is the two strategies of encouraging (punishing) the edges within the layers.
For real-world networks, which usually takes an intermediate structure between the two networks considered here, we can then adopt an appropriate rewarding scheme and weighting the edges and couplings to meet the need.

\subsection{Coupling Contribution}
 \begin{figure*}[!htp]
  \centering{
  \subfloat[I: 4 identical layers.]{\includegraphics[width=0.24\linewidth]{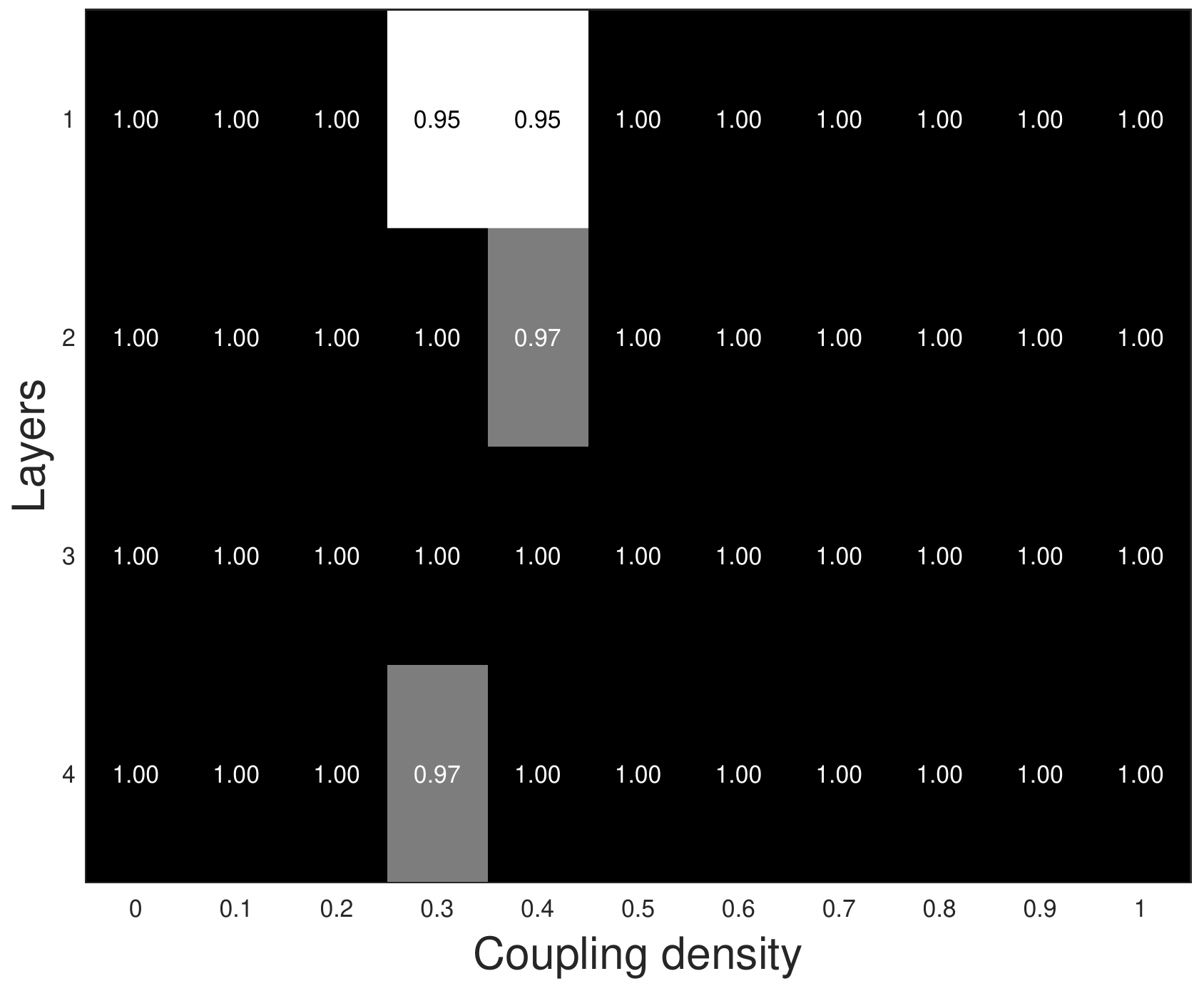}\label{fig:not_considerone}}
  \subfloat[I: 3 identical layers.]{\includegraphics[width=0.24\linewidth]{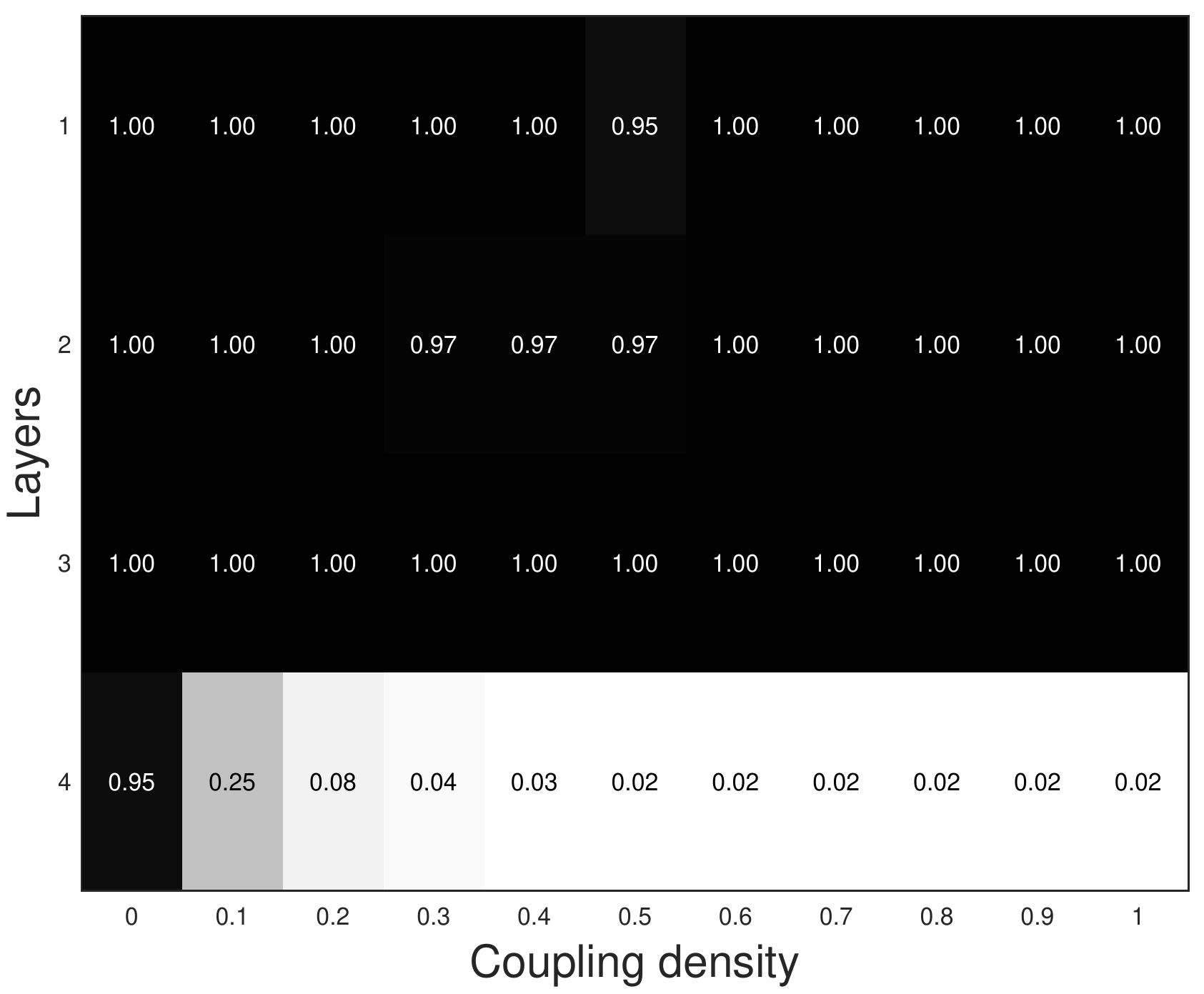}\label{fig:not_considertwo}}
  \subfloat[I: 2 identical layers.]{\includegraphics[width=0.24\linewidth]{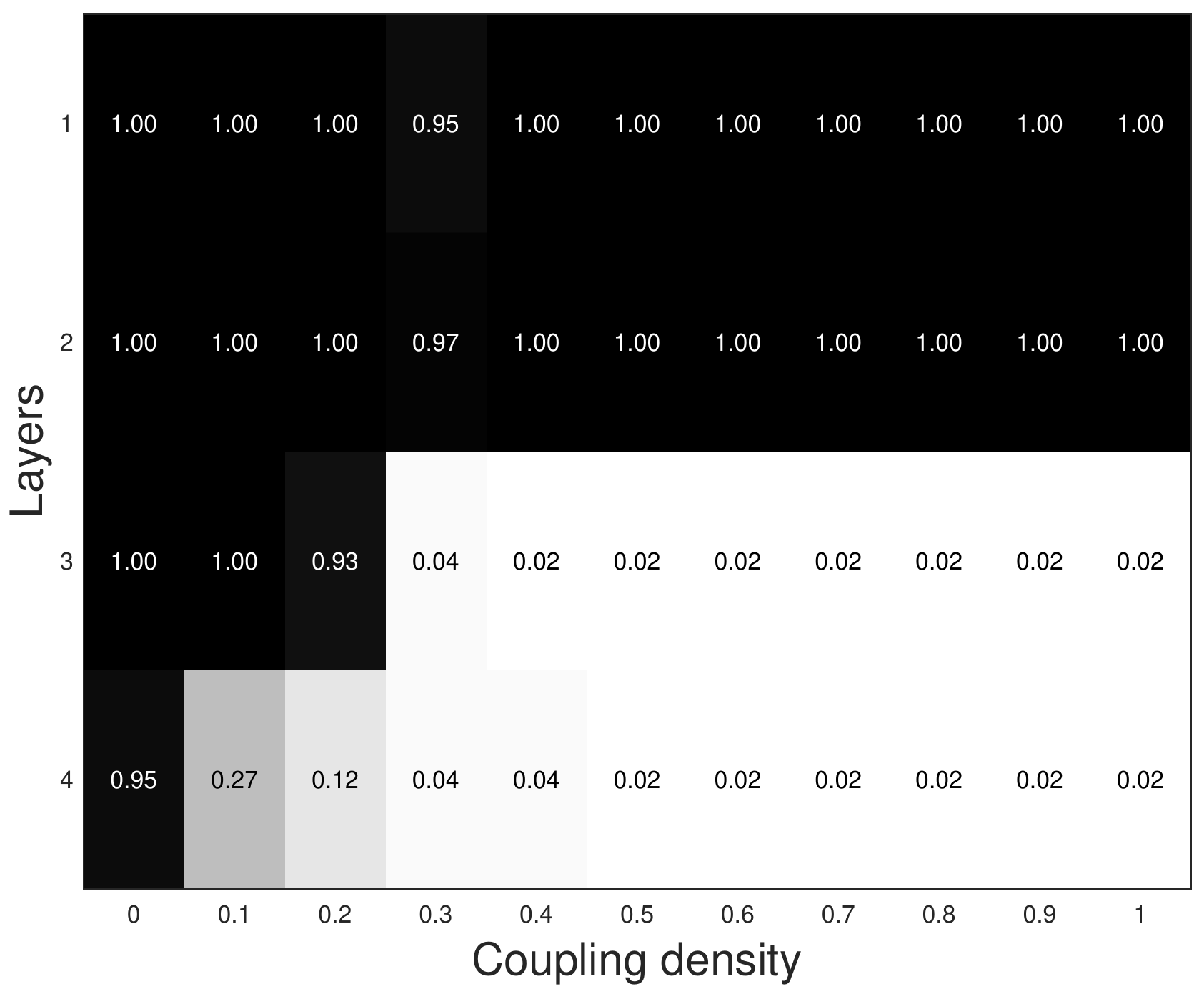}\label{fig:not_considerthree}}
  \subfloat[I: no identical layers.]{\includegraphics[width=0.24\linewidth]{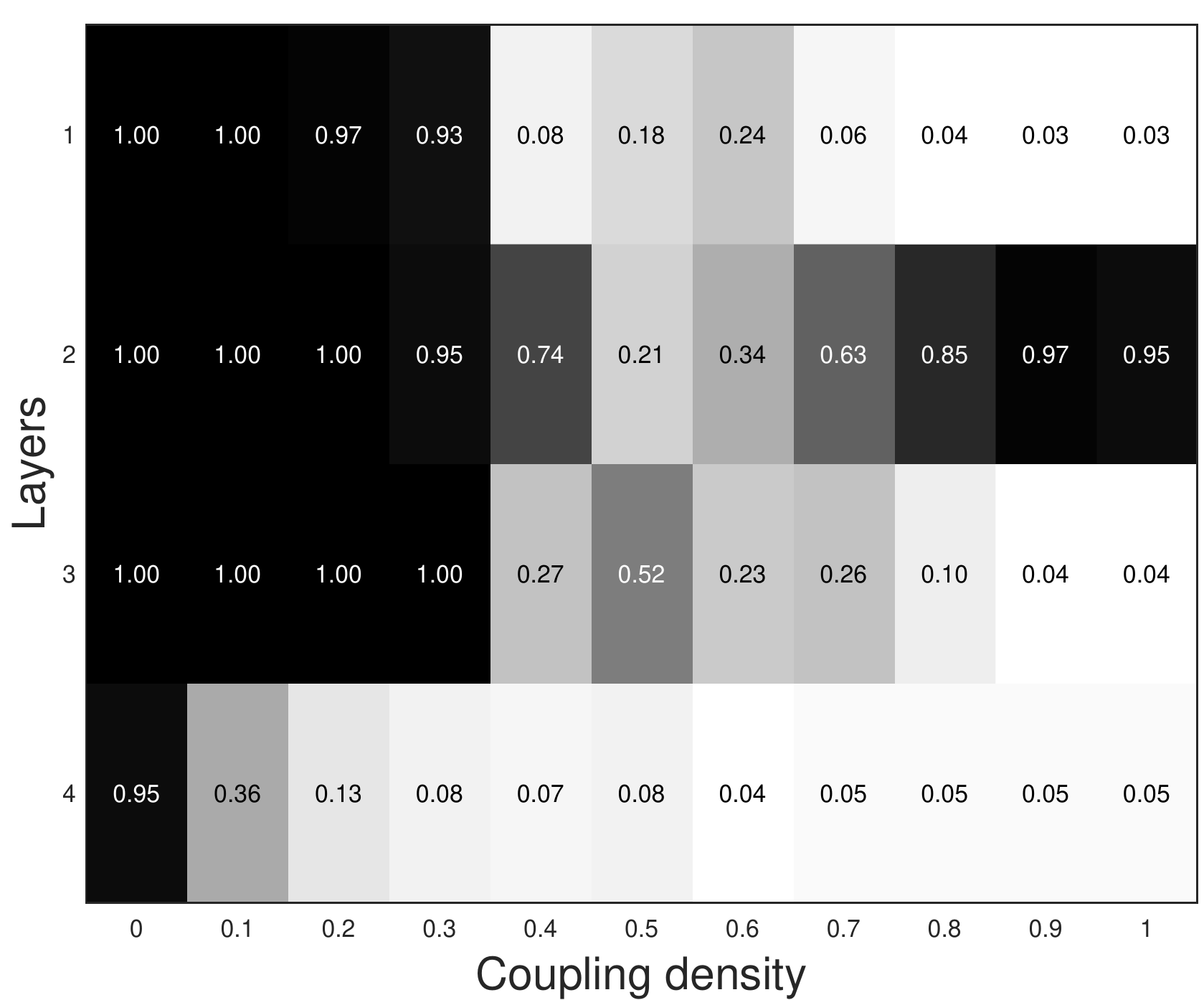}\label{fig:not_considerfour}}

  }
  \centering{
  \subfloat[C: 4 identical layers.]{\includegraphics[width=0.24\linewidth]{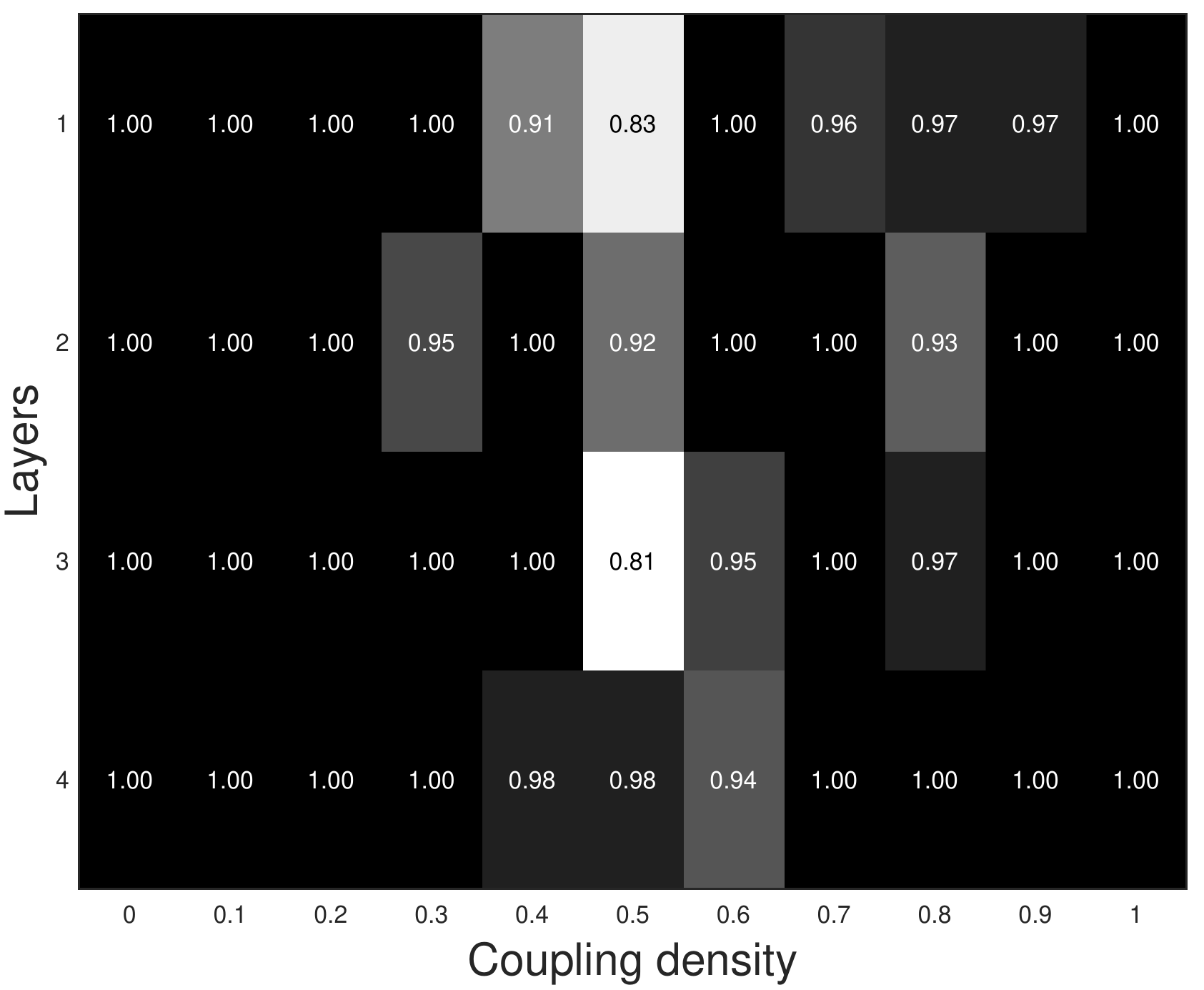}\label{fig:considerone}}
  \subfloat[C: 3 identical layers.]{\includegraphics[width=0.24\linewidth]{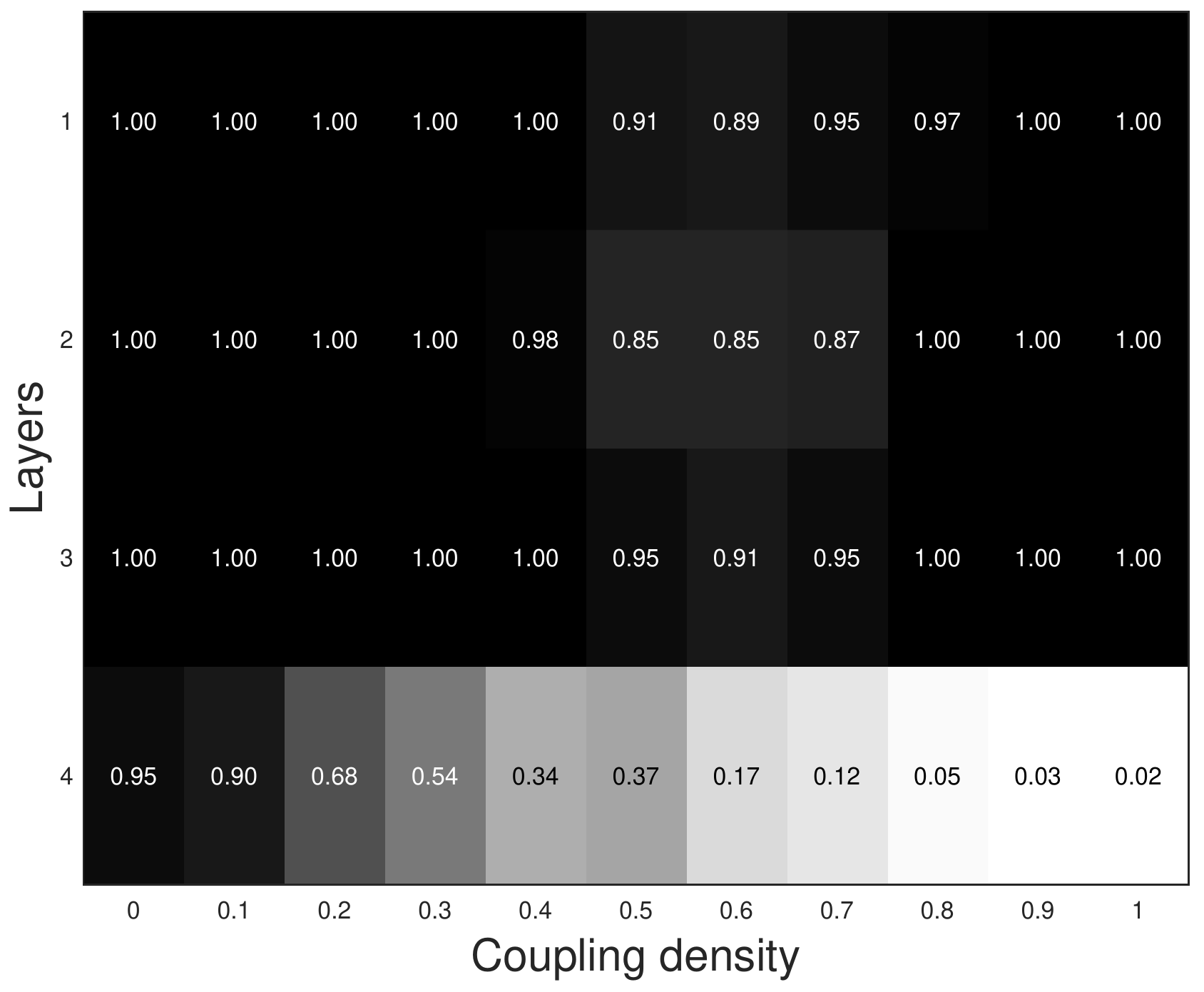}\label{fig:considertwo}}
  \subfloat[C: 2 identical layers.]{\includegraphics[width=0.24\linewidth]{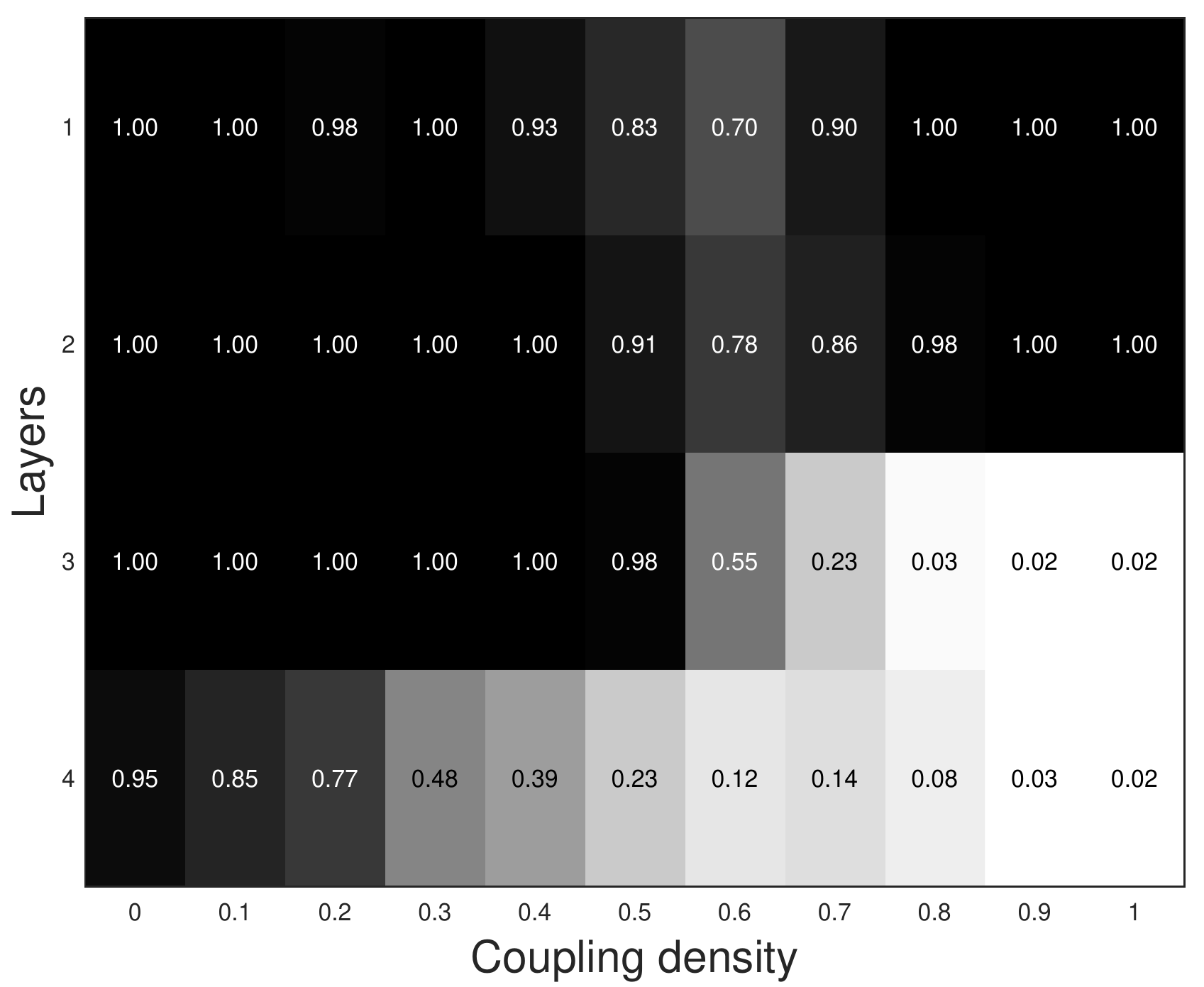}\label{fig:considerthree}}
  \subfloat[C: no identical layers.]{\includegraphics[width=0.24\linewidth]{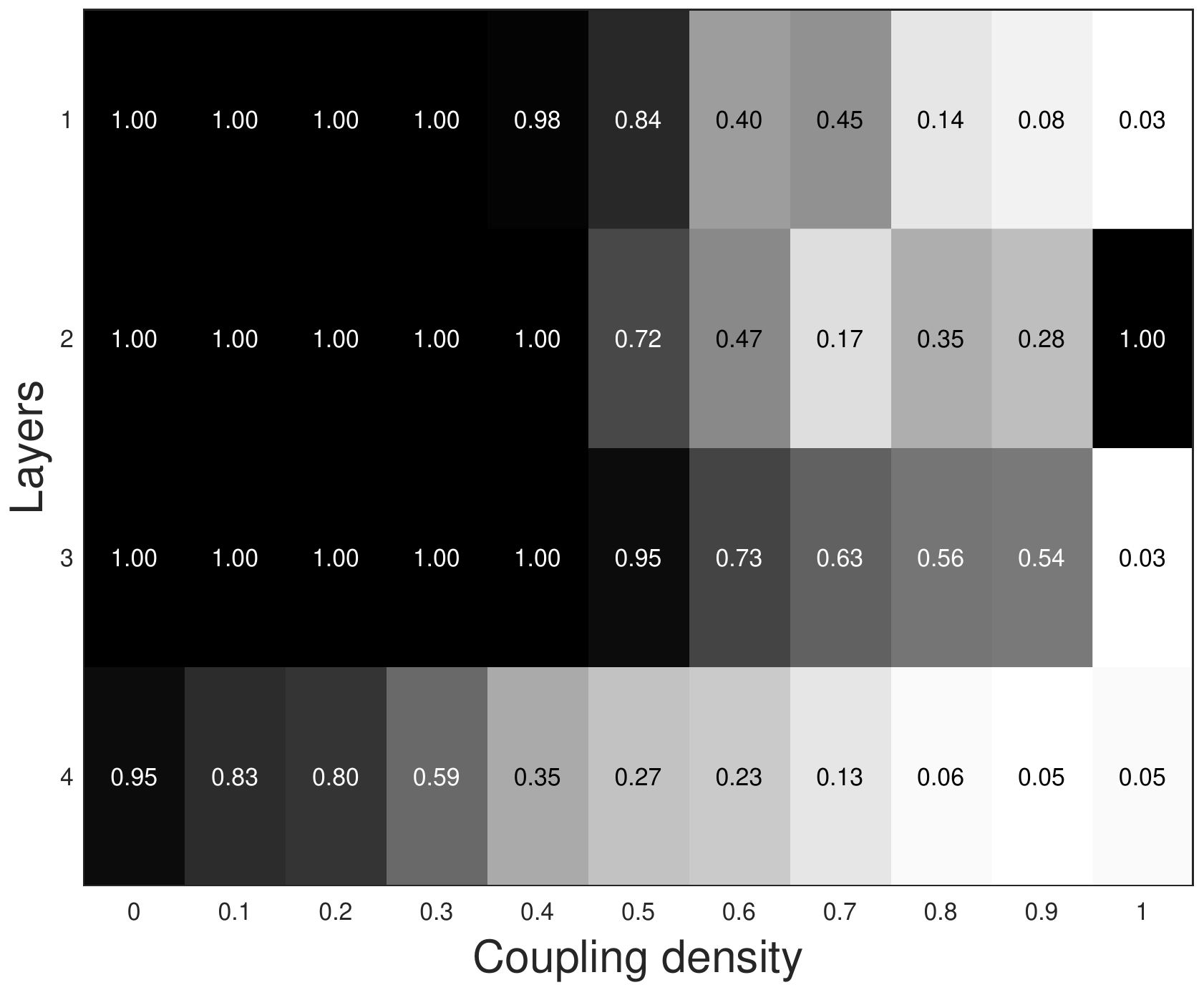}\label{fig:considerfour}}
  }
  \caption{Community detection results in terms of NMI on different networks with different coupling density.
  The figures (a), (b), (c), (d) are obtained by modularity that ignores the non-existing couplings (``I" stands for ``ignore"), and the other four figures are obtained by taking the non-existing couplings into account (``C" stands for ``consider").
  Block $(s, \rho)$ means the NMI value of layer $s$ when the coupling density is $\rho$.
  Deeper color indicates a larger NMI value.
  Different networks consist of different number of identical layers, as stated in corresponding caption.
  }
  \label{fig:differentCouplingDensity}
\end{figure*}

In the multilayer network model, the couplings combine the layers to form a complex network structure.
Thus the contribution of the couplings plays a crucial role in the correlation of the assignments of different layers.
According to Eq. \eqref{eq:choiceOfModularity}, the coupling strength adopted by the multilayer modularity is equivalent to considering an equal contribution of the existing couplings while omitting the non-existing couplings.
In this way, the contribution of the couplings takes $\{0, \varsigma\}$.
If we further take the non-existing couplings into account, i.e. $e_{isr} = f_{isr} = g_{isr} = h_{isr} = \varsigma$ and $\lambda(\{e\}) = \lambda(\{h\}) = -\lambda(\{f\}) = \lambda(\{g\}) = +1$, the contribution of the couplings now takes $\{-\varsigma, \varsigma\}$.
We then apply these two strategies on benchmark networks to see the performance of them when the coupling structure varies.

To study the contribution of couplings to MEMM more comprehensively, we construct benchmark networks with different heterogeneity of layers and different coupling densities.
More specifically, we construct 4 multilayer networks with 0, 2, 3, 4 identical layers in them, respectively.
We first generate 4 LFR benchmark networks with 128 nodes, 4 communities and an average within-layer degree 16, and then duplicate one benchmark network to obtain several copies.
The duplicates are chosen as the identical layers and the rest of the layers are chosen from the other three networks.
We then insert couplings into these 4 networks according to probability $\rho$, which controls the density of the couplings.
In particular, if we obey Definition \ref{def:1} to take $\varsigma = 1$, the two strategies of coupling will differ only when the coupling density $\rho = 1$.
To amplify the difference, we choose $\varsigma = 6$ in this experiment so that the contribution of couplings gets highlighted.
We will discuss about the influence of the coupling strength $\varsigma$ and the resolution parameter $\gamma_s$ in detail later in this section.
 As an indicator of the detection quality, the normalized mutual information (NMI)~\cite{danon:2005comparing, meilua:2007comparing} is calculated between the ground-truth labels and the obtained community assignments.
 NMI measures the similarity between two given vectors and is widely adopted to evaluate the accuracy of a partition, where a higher value indicates a better partition.

We apply the two strategies of coupling contribution to these networks, and report the corresponding NMI for each layer, as demonstrated in \figurename~\ref{fig:differentCouplingDensity}.
The detection results of modularity ignoring and considering the non-existing couplings are shown in the two rows, respectively, and the columns are networks with different heterogeneity of layers.
When the layers are all the same, the two strategies both lead to a good assignment.
As the heterogeneity increases (the number of identical networks decreases), the evaluators show poor results on networks with high coupling density.
This suggests that the couplings will force the layers to make a similar community assignment to the nodes, which result in losing the heterogeneity of layers.
The similar layers will greatly influence the assignments of the other layers.
Moreover, by comparing the NMI of the two strategies, we find that considering the non-existing couplings performs better as the coupling density $\rho$ and the heterogeneity of layers increase, except when the coupling density $\rho$ is around $0.5$.
This confirms that considering the non-existing couplings in MEMM will be more likely to make a balanced assignment when the layers show great heterogeneity and the couplings are dense.
The decline of performance at $\rho = 0.5$ when considering the non-existing couplings in modularity actually arises due to our choice of hyper parameters for the couplings.
We will discuss this gap later in this section.

%Therefore, considering the non-existing couplings will enpower the evaluator a more reliable performance when the coupling structure varies, since it makes fuller use of the information of the network.

\subsection{Parameter Analysis}
The two parameters $\gamma_s$ and $\varsigma$ of multilayer modularity control the behavior of the evaluator.
The resolution limit parameter $\gamma_s$ controls the size of the detected communities, and the coupling strength $\varsigma$ determines the contribution of couplings~\cite{mucha:2010community}.
Nevertheless, if we interpret the parameters of modularity from the perspective of MEMM, varying these parameters signifies violation of Definition \ref{def:1}.
To study their impact, we apply modularity with different $\varsigma=\{0, 0.001, 0.1, 0.5, 1, 10\}$ on a generated network.
This network consists of 11 copies of the Zachary Karate network~\cite{zachary:1977information} as the layers, with corresponding resolution parameter $\gamma_s = \{0, 0.2, 0.4, 0.6, 0.8, 1.0, 1.2, 1.4, 1.6, 1.8, 2.0\}$.
The couplings are inserted with density $\rho = 0.3$ (the largest value that guarantees relatively satisfactory performance as shown in \figurename~\ref{fig:not_considerfour}).
The Zachary Karate network consists of 2 communities in general or 4 communities if considering a more fine-grained partition.

\begin{figure*}[!htp]
  \centering{
  \subfloat[$\varsigma = 0$]{\includegraphics[width=0.32\linewidth]{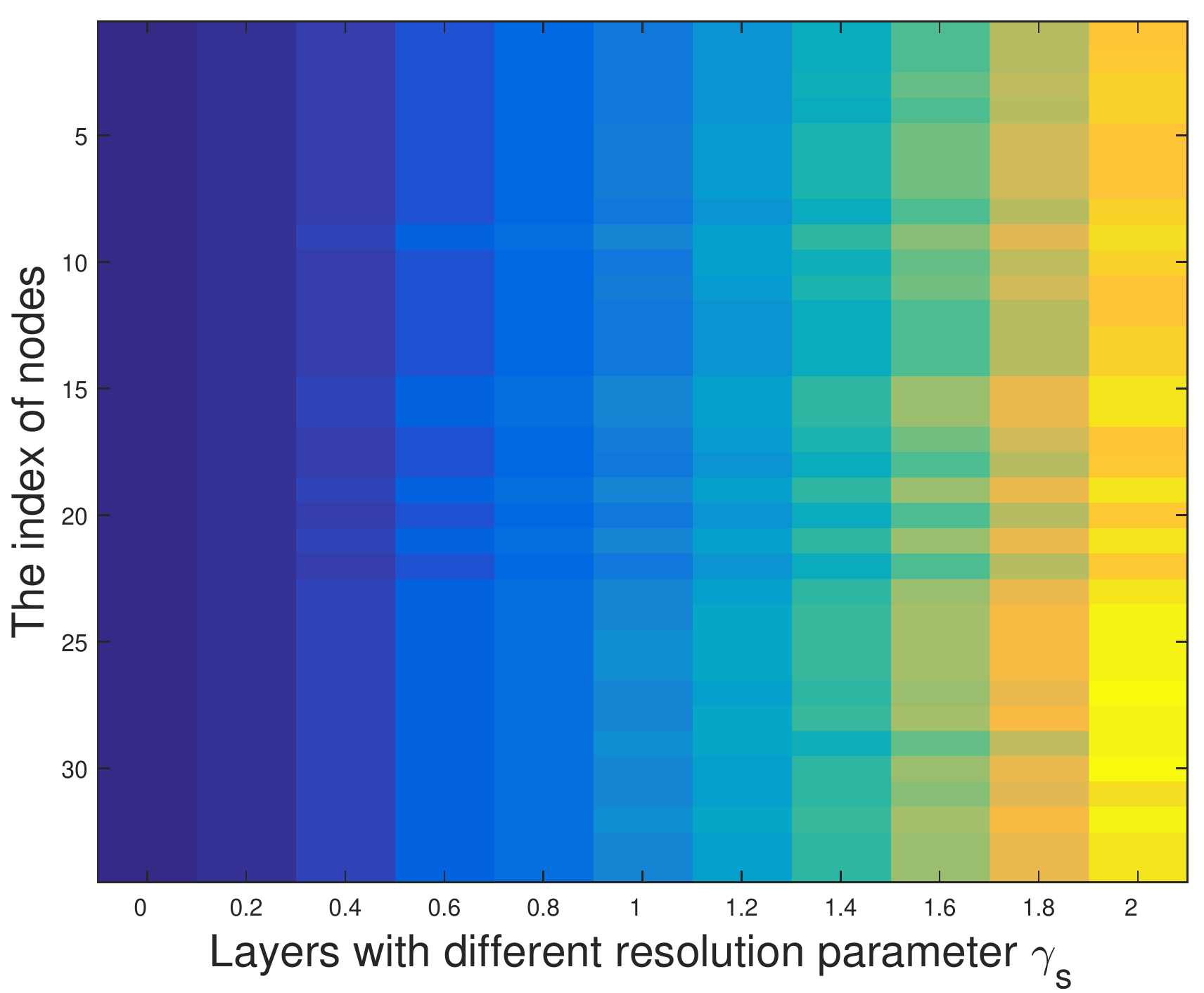}\label{fig:omega_1}}
  \subfloat[$\varsigma = 0.001$]{\includegraphics[width=0.32\linewidth]{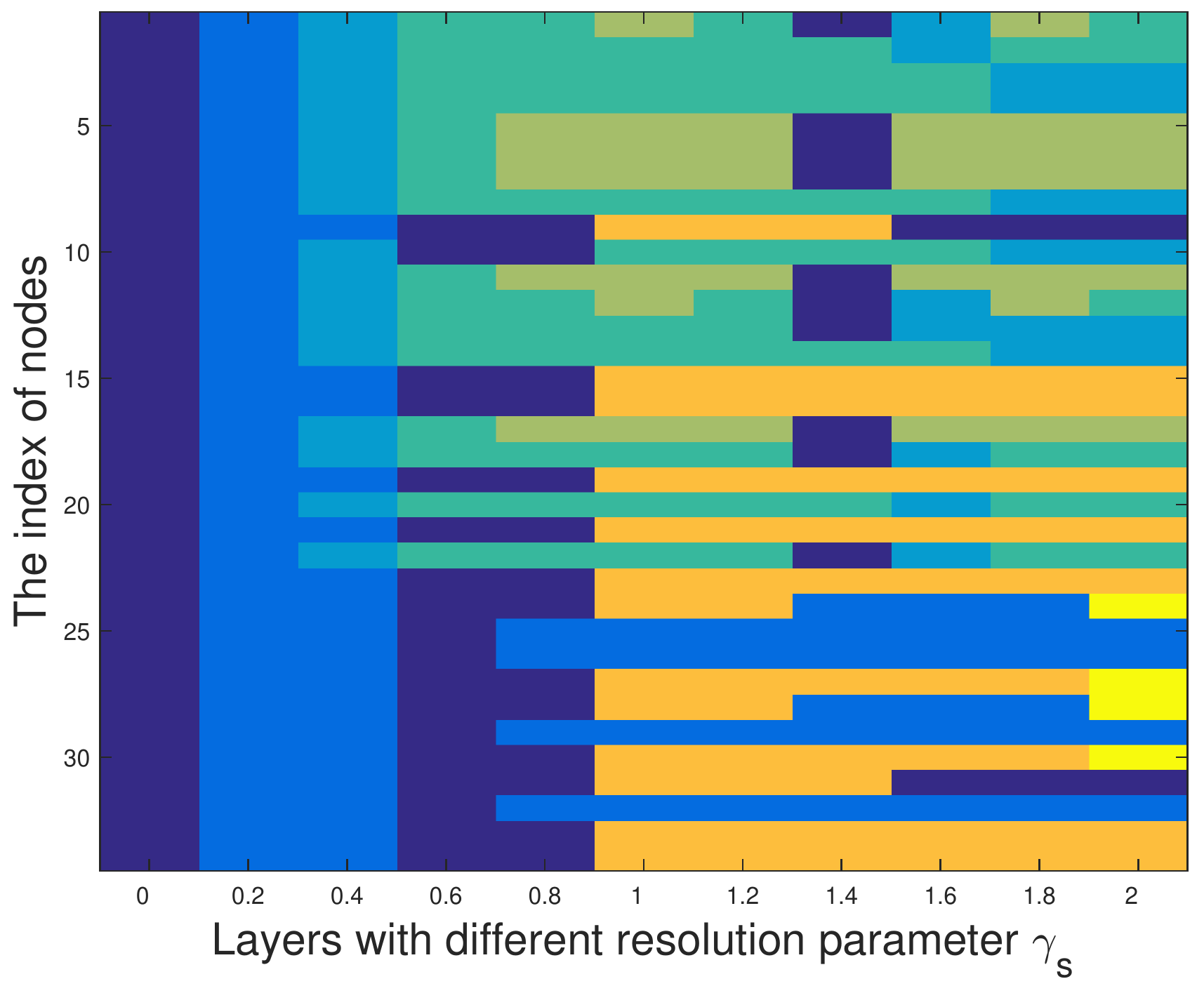}\label{fig:omega_2}}
  \subfloat[$\varsigma = 0.1$]{\includegraphics[width=0.32\linewidth]{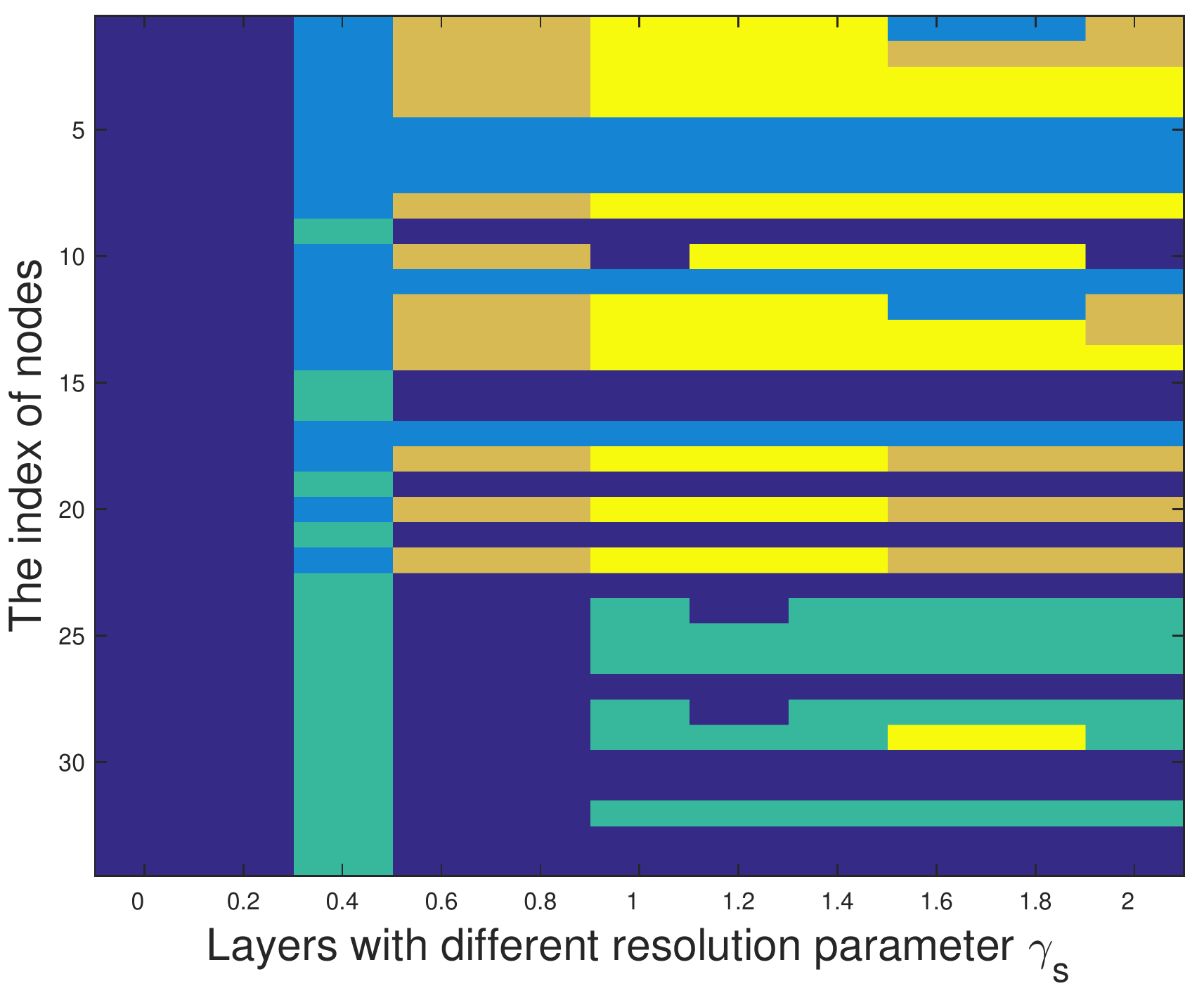}\label{fig:omega_3}}
  }
  \centering{
  \subfloat[$\varsigma = 0.5$]{\includegraphics[width=0.32\linewidth]{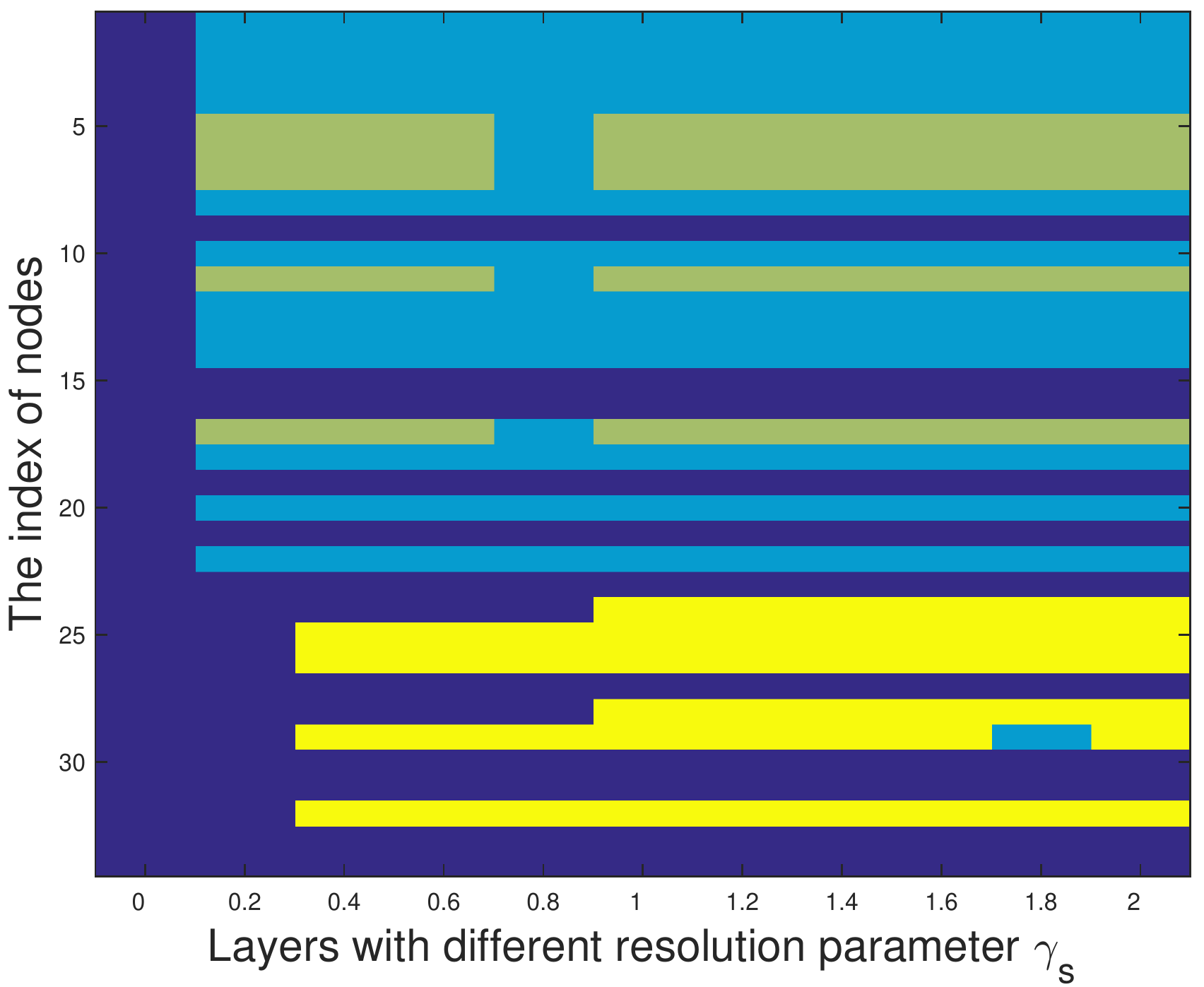}\label{fig:omega_4}}
  \subfloat[$\varsigma = 1$]{\includegraphics[width=0.32\linewidth]{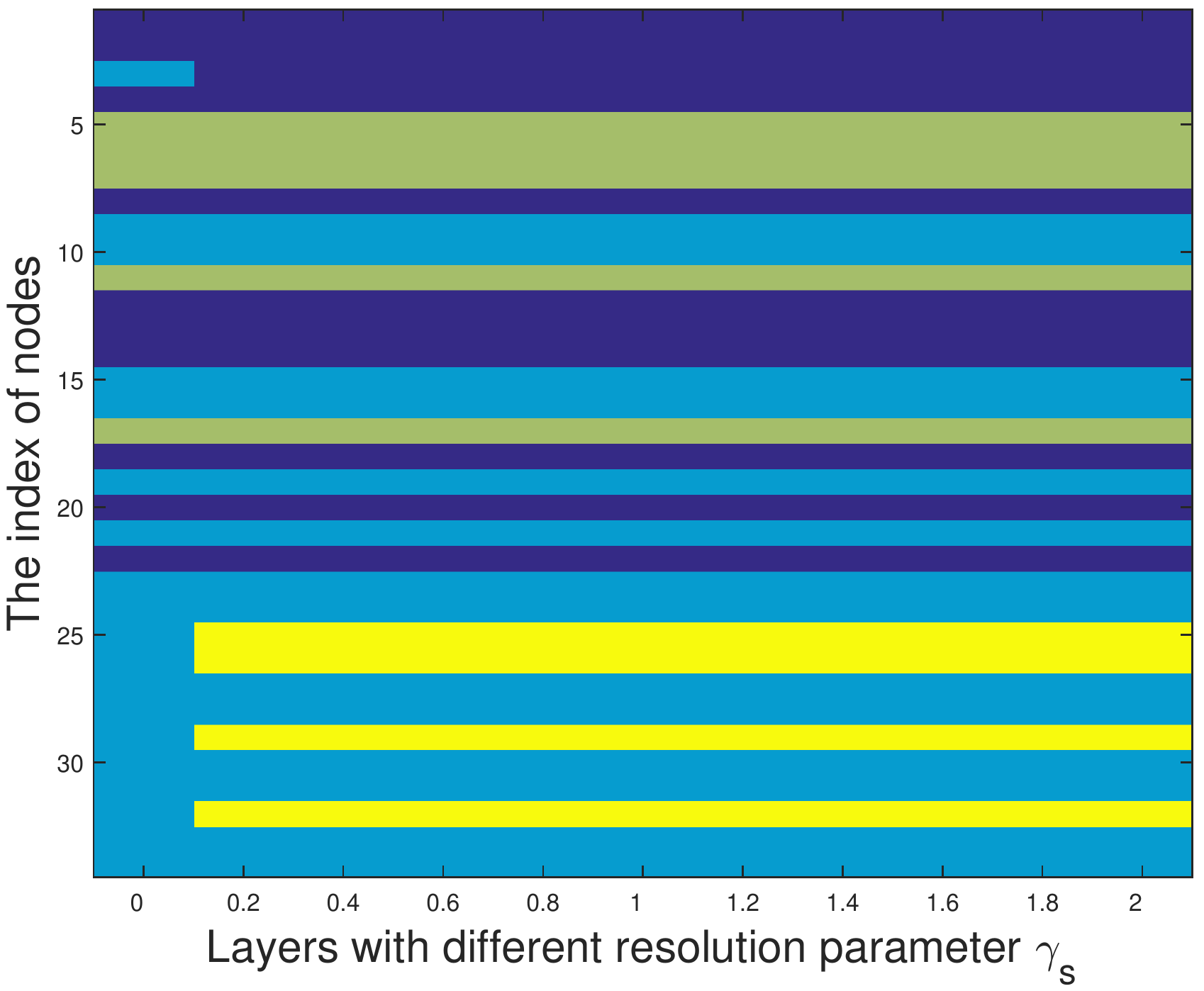}\label{fig:omega_5}}
  \subfloat[$\varsigma = 10$]{\includegraphics[width=0.32\linewidth]{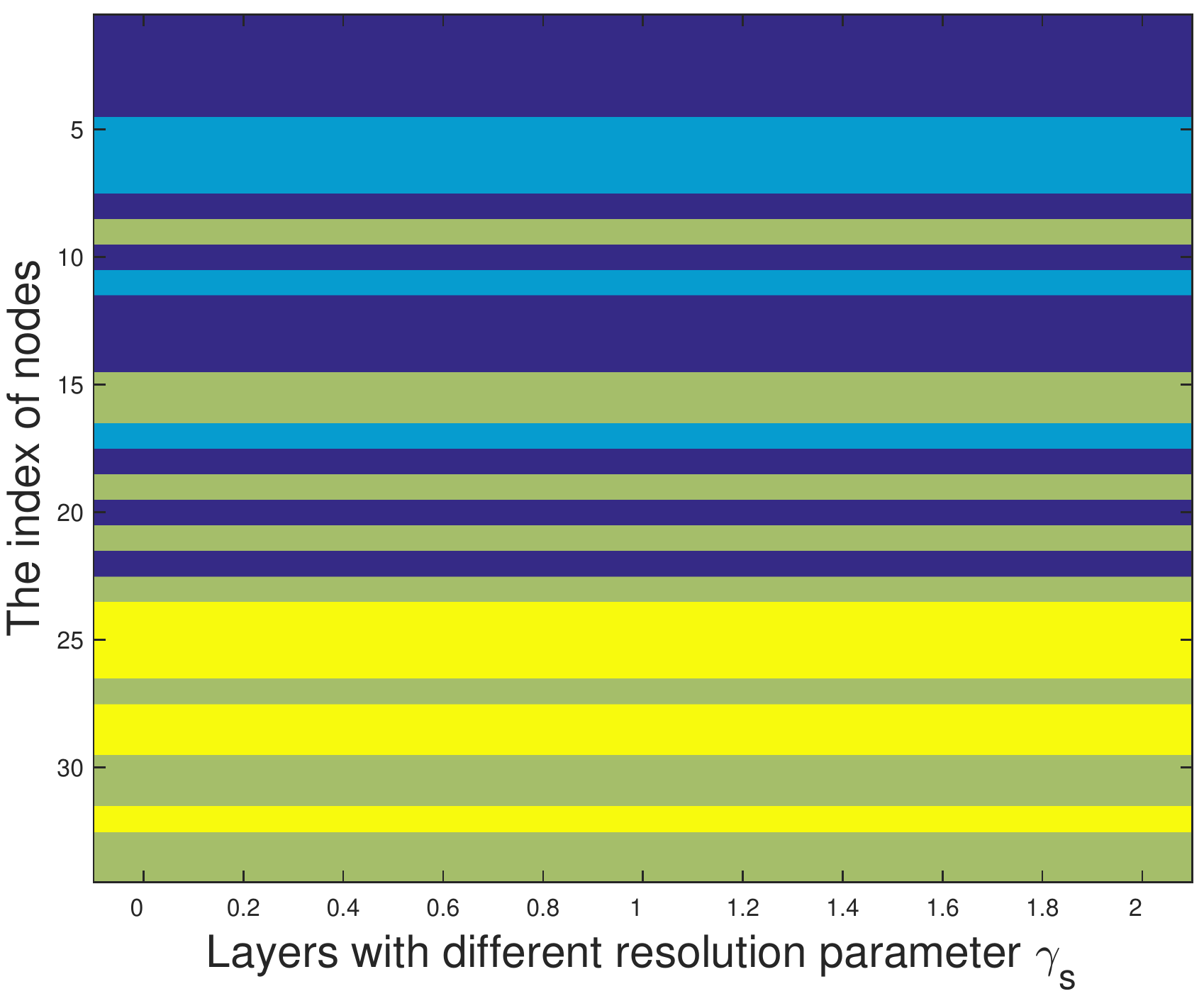}\label{fig:omega_6}}
  }
  \caption{Community assignments for different resolution parameters $\gamma_s$ and coupling strength $\varsigma$.
  Different colors indicate different community labels.}
  \label{fig:parameterAnalysis}
\end{figure*}
The community assignments are shown in \figurename~\ref{fig:parameterAnalysis}.
We see that, when $\varsigma = 0$, there is no community across two layers, and as $\gamma_s$ increases, the partition becomes more fine-grained.
However, the community assignments in different layers are similar but not identical, which implies there must be some misclassifications.
As $\varsigma$ increases, the communities tend to stretch across the layers and the assignments of a node and its copies in other layers become more unanimous, while there are still some misclassifications.
The resolution difference is completely overpowered by the couplings when taking a large coupling strength $\varsigma$.

We can then conclude that, the resolution parameter $\gamma_s$ controls the tendency of the splitting and the coupling strength parameter $\varsigma$ controls the consistency of the community assignment between layers.
Too large or too small $\gamma_s$ will cause misclassification, which can be fixed, however, by the view couplings.
Meanwhile, too small $\varsigma$ will lead to the isolation between layers.
When there are noises in the network data, the result can be poor for those layers that suffer severe interference of noise since the coupling information has not been fully utilized.
Nevertheless, the peculiarity of each view will be damaged by large $\varsigma$ (as shown in \figurename~\ref{fig:omega_6}).
Therefore, it is more reliable to choose the parameters suggested by Definition \ref{def:1}.

\subsection{Discussion}
In this section, we discuss the decline of modularity's performance when considering the non-existing couplings at around $\rho = 0.5$.
In fact, if we denote the number of the internal and external existing couplings as $\mathcal{C}_{I}$ and $\mathcal{C}_{E}$, and the non-existing couplings as $\mathcal{C}_{\bar{I}}$ and $\mathcal{C}_{\bar{E}}$, we can rewrite the contribution of the couplings (considering the non-existing couplings) as
\begin{equation}
\label{eq:couplingContributionDecomposition}
\begin{aligned}
  \mathcal{M}_{coupling} &= \varsigma(\mathcal{C}_{I}-\mathcal{C}_{\bar{I}}-\mathcal{C}_{E}+\mathcal{C}_{\bar{E}})\\
  &=\varsigma\big[2(\mathcal{C}_{I}+\mathcal{C}_{\bar{E}}) - M\big],
\end{aligned}
\end{equation}
where $M = \mathcal{C}_{I}+\mathcal{C}_{\bar{I}}+\mathcal{C}_{E}+\mathcal{C}_{\bar{E}}=l(l-1)N$ is the number of all possible couplings in the network, $N$ is the number of nodes within a layer and $l$ is the number of layers.
Here, the number of couplings of each type $\mathcal{C}_{x}$ depends on the coupling density $\rho$ and the current community assignment.

We can see that the contribution of the couplings depends on the number of couplings present in the communities and absent between different communities, fixing the coupling strength.
As $\rho$ increases, $\mathcal{M}_{coupling}$ goes to $\varsigma(2M_{I}-M)$, where $M_{I} = \mathcal{C}_{I}\mid_{\rho=1} = \mathcal{C}_{\bar{I}}\mid_{\rho=0}$.
 This indicates that the contribution is governed by the total amount of internal couplings.
Similarly, if $\rho \rightarrow 0$ the contribution will be governed by the external non-existing couplings with maximum $\varsigma(2M_{E}-M)$, where $M_{E} = \mathcal{C}_{E}\mid_{\rho=1} = \mathcal{C}_{\bar{E}}\mid_{\rho=0}$.
When $\rho$ varies, we have
\begin{equation}
\label{eq:couplingContributionDecomposition2}
\begin{aligned}
  \frac{\mathcal{M}_{coupling}}{\varsigma} &= 2\big[\rho M_I + (1-\rho)M_E\big] - M\\
  &= 2\big[\rho (M-M_E) + (1-\rho)M_E\big] - M\\
  &= (2\rho-1)(M_E-M_I),
\end{aligned}
\end{equation}
given the community assignment.
The quantity $M_E$ and $M_I$ are functions of the community assignment.
Thus the contribution of the couplings is composed of the strength parameter $\varsigma$, the current community assignment and the coupling strength.
Eq. \eqref{eq:couplingContributionDecomposition2} suggests that the contribution of couplings is expected to be positive if the coupling density is high and there are more internal couplings $M_I$ than external couplings $M_E$, or low $\rho$ with more external couplings.
In particular, the contribution is expected to vanish when the coupling density $\rho$ goes to $0.5$.
At this time, the Louvain method will generate a less optimal assignment owing to the heuristically merge.
The global contribution of the couplings is expected to be zero, but when the Louvain method attempts to merge two communities to locally increase the modularity, it will take the couplings into consideration.
Thus the error is accumulated during the iteration and finally results in a less satisfactory result.

In contrast, if we omit the terms of non-existing couplings, we obtain a different representation of the contribution:
\begin{equation}
  \begin{aligned}
    \frac{\mathcal{M}_{coupling}}{\varsigma} &= \mathcal{C}_I - \mathcal{C}_E\\
    &=\rho (M_I - M_E).\\
  \end{aligned}
\end{equation}
It will not suffer the problem of contribution vanish as it is a linear function of $\rho$.
However, the influence of couplings increases as $\rho$ increases, to force the layers to adopt a similar assignment.
As a result, the heterogeneity of layers is erased.
\section{Conclusion}
\label{sec:conclusion}
In this paper, we presented the multilayer edge mixture model which regards the community structure as the combination of the edge (coupling) contributions.
We studied how the multilayer edge mixture model qualifies the community structure by generating an evaluator according to the definition of the community, and discussed how to choose a discriminative hyper parameters set for balancing the contributions.
Multilayer modularity and stochastic blockmodels have been derived from the proposed model as an example.
We also discussed how to decompose a community structure evaluator with specific forms to the multilayer edge mixture model, which reveals the preference of the evaluator on the edges and couplings.

We compared the performance of the modularity with its modified form on two different network types, to show that the rewarding scheme, specified by the $\lambda$ function in the proposed model, greatly influences the detected community structure.
The application of modularity on networks with two coupling contribution strategies shows the impact of ignoring some types of couplings in the proposed model on the detection accuracy as the coupling density and layer heterogeneity vary.
We analyzed the coupling strength and resolution parameter in modularity, interpreted as the derived evaluator of the proposed model, to show that extreme values of the parameters will affect the detection result.
The parameters are recommended to take values that obeys the proposed discriminative definition.

The multilayer edge mixture model is able to derive other desired evaluators, based on the practical definition of the community, and help interpret the nature of the evaluators.
Therefore, it is helpful both in deriving new quality functions for practical need and studying the existing community evaluators.

%\section*{Acknowledgements}
%We'd like to thank Sun Yat-sen Memorial Hospital for providing EEG data.

\bibliographystyle{apalike}

\bibliography{MEMM}
\end{document}